\documentclass[aps,prc,reprint,superscriptaddress,nofootinbib,floatfix,showpacs]{revtex4-1}
\pdfoutput=1

\usepackage{epsfig}
\usepackage{dcolumn}
\usepackage{bm}
\usepackage{amssymb}
\usepackage{amsmath}
\usepackage{hyperref}
\usepackage{xcolor}

\newcommand{\oxygen}{{}^{16}\mathrm{O}}
\newcommand{\argon}{{}^{40}\mathrm{Ar}}

\begin{document}
\title{Shrinking the Quark Gluon Plasma}

\author{M. Sievert}
\affiliation{Department of Physics and Astronomy, Rutgers University, Piscataway, NJ USA 08854}
\author{J. Noronha-Hostler}
\affiliation{Department of Physics and Astronomy, Rutgers University, Piscataway, NJ USA 08854}

\date{\today}
\begin{abstract}
In recent years the understanding on the limits of the smallest possible droplet of the Quark Gluon Plasma has been called into question.  Experimental results from both the Large Hadron Collider and the Relativistic Heavy Ion Collider have provided hints that the Quark Gluon Plasma may be produced in systems as small as that formed in pPb  or dAu collisions.  Yet alternative explanations still exist from correlations arising from quarks and gluons in a color glass condensate picture.  In order to resolve these two scenarios, a system size scan has been proposed at the Large Hadron Collider for collisions of ArAr and OO.  Here we make predictions for a possible future run of ArAr and OO collisions at the Large Hadron Collider and study the system size dependence of a variety of flow observables.  We find that linear response (from the initial conditions to the final flow harmonics) becomes more dominant in smaller systems whereas linear+cubic response can accurately predict multi-particle cumulants for a wide range of centralities in large systems. 
\end{abstract}

\maketitle

\section{Introduction}

Since the early 2000's the Quark Gluon Plasma (QGP) has been produced in relativistic heavy ion collisions at the Large Hadron Collider (LHC) and the Relativistic Heavy-Ion Collider (RHIC) in symmetric ion on ion (AA) collisions.  One of the pillars of the detection of the QGP is long range multiparticle correlations in the low transverse momentum ($p_T$) sector known as collective flow that can be very well-reproduced and predicted using event-by-event relativistic viscous hydrodynamics. Theoretical calculations \cite{Niemi:2015voa,Noronha-Hostler:2015uye,Eskola:2017bup,Giacalone:2017dud} in AA collision have been accurate enough to predict enhancements to the flow harmonics at the level of a few percentage points when the center of mass beam energy is increased.  

In 2013 the three heavy-ion LHC experiments that measure collective flow (ATLAS \cite{Chatrchyan:2013nka,Aaboud:2017acw,Aaboud:2017blb,Aad:2013fja}, CMS \cite{Sirunyan:2018toe,Chatrchyan:2013nka,Khachatryan:2014jra,Khachatryan:2015waa,Khachatryan:2015oea,Sirunyan:2017uyl}, and 
ALICE \cite{ABELEV:2013wsa,Abelev:2014mda})  all released results indicating that asymmetric pPb collisions have signatures of collective flow and even pp collisions may also have collective flow signatures (although the latter is significantly less certain).  While the experimental data can now be quantitatively reproduced by hydrodynamics \cite{Bozek:2011if,Bozek:2012gr,Bozek:2013ska,Bozek:2013uha,Kozlov:2014fqa,Zhou:2015iba,Zhao:2017rgg,Mantysaari:2017cni,Weller:2017tsr,Zhao:2017rgg} with certain caveats (such as using eccentric protons for the initial conditions \cite{Mantysaari:2017cni}), alternative explanations for these long range correlations have emerged  \cite{Greif:2017bnr,Schenke:2016lrs,Mantysaari:2016ykx,Albacete:2017ajt}. The question of if these collective signatures can be attributed to a nearly perfect fluid stage of QGP or if they arise purely from quark gluon correlations in the initial stage via a  ``color glass condensate" (CGC) picture \cite{Kovchegov:2012mbw,Iancu:2001ad,McLerran:1993ni,McLerran:1993ka,McLerran:1994vd,Iancu:2000hn} is one of the most hotly contested in the field.

Experimentalists at the RHIC PHENIX detctor \cite{Aidala:2018mcw,Adare:2018toe} recently entered the debate with a beam energy scan of ${}^3\mathrm{He} \mathrm{Au}$ and $\mathrm{d} \mathrm{Au}$ collisions that also clearly showed signs of collective behavior.  The lower beam energies at RHIC produce an even smaller and cooler QGP so these results imply that if the hydrodynamic picture is the answer then a QGP can be produced even in tiny ``cool" systems  \cite{Nagle:2013lja}. However, recent improvements in the CGC also claim that they can reproduce these results \cite{Mace:2018vwq}, which has been contested by experimentalists \cite{Nagle:2018ybc, PHENIX:2018lia}.

At the LHC ${}^{129}\mathrm{Xe}$ collisions were used to study a slightly smaller system size, which had the quite unexpected result of measuring a deformed ${}^{129}\mathrm{Xe}$ nucleus in ultra central collisions \cite{Giacalone:2017dud,Acharya:2018ihu,Acharya:2018eaq,CMS:2018jmx,ATLAS:2018iom}. Beyond the deformation, however, it appears that most theoretical predictions quantitatively hold up to experimental data \cite{Giacalone:2018cuy,Eskola:2017bup} such that the hydrodynamical scaling between PbPb and XeXe collisions is well understood.  The question still remains how this would hold up in smaller symmetric ion collisions.

The intrinsic question of the origin of these long-range collective multiparticle behavior is intriguing but there are still other physical properties that can be learned from studying small systems. For instance, flow fluctuations have been shown to play a discriminating role in the preferred initial conditions used in heavy-ion collisions \cite{Giacalone:2017uqx} and they demonstrate differences in the sensitivity to smoothing out small scale structure \cite{Giacalone:2017uqx} in pPb collisions. Additionally, strangeness enhancement (another signature of the QGP) appears to gradually turn off as the system size is decreased \cite{ALICE:2017jyt}. Finally, as of yet the third main signature of the QGP, jet quenching, has not been measured in small systems. 

One proposed solution to resolve the above questions about the smallest droplet of the QGP is to run polarized ion beams \cite{Bozek:2018xzy}, which is certainly an appealing suggestion but may take years to create the proper infrastructure to do so.  Thus, in this paper we make predictions for the proposed LHC run ${}^{40}\mathrm{Ar}$ and ${}^{16}\mathrm{O}$ collisions \cite{Citron:2018lsq} (see also \cite{Lim:2018huo,Bruce:2018yzs}) in order to gradually see the signatures of the QGP scale with the system size.  These intermediate ion collisions could help shed light on the scaling of hydrodynamic observables (as explored in this paper) but may also provide relevant information about jet quenching in intermediate systems as well as strangeness enhancement.  

In this paper we are able to draw a number of conclusions based on the scaling of a variety of flow observables with systems size.  For instance, for a fixed number of participants, initial conditions with a smaller radius have a strong linear mapping onto the final flow harmonics whereas larger systems experience both linear and cubic response, which can quite accurately predict the final flow fluctuations of $v_2\{4\}/v_2\{2\}$.  Furthermore, we investigated a variety of flow observables and found that they either fall onto a universal curve when plotted versus the number of participants or exhibit a clear hierarchy depending on the system size. 

This paper is organized as follows: In Section \ref{sec:IC} the parameters for the initial conditions implemented using TRENTO are discussed.  In Section \ref{sec:hydro} the hydrodynamical setup using v-USPhydro is discussed. In Section \ref{sec:linear} the validity of a linear mapping between eccentricities into the final state is tested as one shrinks the system system.  In Section \ref{sec:results} the predictions for a variety of flow observables (multiparticle cumulants, symmetric cummulants, mean transverse momentum, event plane correlations) are shown scaled by either the centrality or Npart. Finally, the Conclusions and Outlook are in Section \ref{sec:conclu}.

\section{Initial Conditions}\label{sec:IC}

In this paper we use TRENTO initial conditions tuned to reproduce IP-GLASMA eccentricities such that p=0, k=1.4, and $\sigma$=0.51 \cite{Moreland:2014oya,Bernhard:2016tnd}. 

\subsection{Parameterizations of Nuclear Geometry}

Heavy-ion collisions involving spherical nuclei are well described by sampling the distribution of nucleons from a two-parameter Woods-Saxon density distribution in the nuclear rest frame, written in spherical coordinates as:
\begin{align} \label{e:2PF}
\rho(r,\theta,\phi) = \rho_0 \left[ 1 + \exp\left(\frac{r - R}{a}\right)\right]^{-1}
\end{align}
with $\rho_0$ the nuclear saturation density, $R$ a measure of the gluonic radius of the nucleus, and $a$ the surface diffusion parameter.  For some nuclei a three-parameter generalization \cite{DeJager:1987qc} of the nuclear density has been extracted instead of (or in addition to) the the standard Woods-Saxon distribution \eqref{e:2PF}.  This three-parameter fit modifies the radial density distribution somewhat:
\begin{align} \label{e:3PF}
\rho(r,\theta,\phi) = \rho_0 \left( 1 + w \frac{r^2}{R^2} \right)
\left[ 1 + \exp\left(\frac{r - R}{a}\right)\right]^{-1} .
\end{align}
For nuclei such as ${}^{208}Pb$, a ``doubly magic'' nucleus in the nuclear shell model, these spherically symmetric densities give a good description of elliptic flow at the LHC.  

For some ``non-magic'' nuclei, however, nonzero electric quadrupole moments have been measured \cite{Stone:2005rzh}, indicating a potentially significant quadrupole deformation in the nuclear geometry.  For such nuclei a non-spherical parameterization can be included by generalizing \eqref{e:2PF} or \eqref{e:3PF} to allow for an angular modulation of the radius (see e.g. \cite{Hamamoto:2011wn}):
\begin{align}
R(\theta) = R \Big(1 + \beta_2 Y_{20} (\theta) + \beta_4 Y_{40} (\theta) + \cdots \Big) .
\end{align}
These non-spherical deformations have been shown to produce a significant enhancement of anisotropic flow in ultra-central collisions of non-spherical nuclei, such as ${}^{129}\mathrm{Xe}$ \cite{Giacalone:2017dud,Acharya:2018ihu,Acharya:2018eaq,CMS:2018jmx,ATLAS:2018iom} and ${}^{238}\mathrm{U}$ \cite{Adamczyk:2015obl, Rybczynski:2012av, Moreland:2014oya, Goldschmidt:2015qya, Schenke:2014tga}.  

\begin{table}
\begin{tabular}{|c|c|cccc|} 
	\hline
			& Parameterization	& $R$ (fm)	& $a$ (fm)	& $w$ (fm)	& $\beta_2$ 	\\
	\hline
	${}^{16}$O & 3pF	& 2.608		& 0.513		& -0.051	& 0 \\
	${}^{40}$Ar & 2pF	& 3.53		& 0.542		& 			& 0 \\
			& 3pF	& 3.73		& 0.62		& -0.19		& 0 \\
	\hline
\end{tabular}
\caption{Parameters for the nuclear density distributions used in the initial conditions.  In all cases we use $\beta_4 = 0$.}
\label{t:params}
\end{table}
The parameters used in our initial conditions are given in Table~\ref{t:params}.  For $\oxygen$, only the three-parameter fit \eqref{e:3PF} is available, while both two- and three-parameter fits are available for $\argon$.  We have studied the effects of both parameterizations for $\argon$ and found no significant differences for our present purposes; accordingly we use here the three-parameter fit for both $\oxygen$ and $\argon$ for consistency.  

Being a doubly magic nucleus, $\oxygen$ is taken to be spherically symmetric with $\beta_2=0$.  For the case of $\argon$, we also assume spherical symmetry $\beta_2 =0$ in the present work, but it is interesting to note that a small nonzero quadrupole moment has been measured \cite{PhysRevLett.24.903} using nuclear scattering techniques.  This measurement does not apply to the nuclear ground state, but rather to an excitation $1.46~\mathrm{MeV}$ above the ground state.  Since it is unclear to what extent such a deformation may be present in the ground state, it may interesting to explore the potential effects of spherical versus deformed geometry for $\argon$, which we leave for future work.  Such studies of the role of nuclear deformations in heavy-ion collisions have the potential to provide new constraints on the nuclear shape parameters, complementing the limited extractions available from low-energy scattering experiments \cite{DeJager:1987qc, Stone:2005rzh}.

In the previous theoretical predictions for XeXe collisions at $\sqrt{s_{NN}}=5.44$ TeV compared to PbPb collisions at  $\sqrt{s_{NN}}=5.02$, an inelastic cross-section, $\sigma_{NN}$, was set at $\sigma_{NN}=70$ mb for  $\sqrt{s_{NN}}=5.02$ and $\sqrt{s_{NN}}=5.44$ TeV with the assumption that there was almost no difference in the cross-section at those energies. Previous experimental data at the LHC still has large uncertainties for the energy of $\sqrt{s_{NN}}=7$ TeV  where if all error bars are taken seriously across all LHC experiments $\sigma_{NN}$ could be anywhere in the range of $\sigma_{NN}=59.6-77.2$ mb   taken from ATLAS \cite{Aad:2011eu}, CMS \cite{CMS:2011xpa},  and ALICE \cite{Abelev:2012sea}.  Here we assumed there would be a small increase in $\sigma_{NN}$ with beam energy such that at $\sqrt{s_{NN}}=5.85$ TeV  $\sigma_{NN}= 71$ mb and $\sqrt{s_{NN}}=6.5$ TeV $\sigma_{NN}= 72.5$ mb.  We then checked if this modest increase in $\sqrt{s_{NN}}$ played any role in the eccentricities comparing OO collision using $\sigma_{NN}= 72.5$ mb and $\sigma_{NN}= 70$ mb. We found no discernible difference in the eccentricities so we do no expect the influence of the error bars in $\sigma_{NN}$ to influence our results shown here. 

\subsection{Eccentricities}

In relativistic heavy ion collisions, the initial conditions are quantified by the eccentricity vectors $\mathcal{E}=\varepsilon_n e^{i\phi_n}$ where $\varepsilon_n$ is the magnitude and $\phi_n$ is the angle.  The eccentricities can be calculated using:
\begin{equation}
\varepsilon_n=\frac{\int r^n e^{in\phi}s(r,\phi) r drd\phi}{\int r^n s(r,\phi) r drd\phi}
\end{equation}
where the position is taken in reference from the center of mass. In this paper we limit ourselves to $n=2, 3$ because other measured harmonics $n\geq 4$ and $n=1$ involve nonlinear mode mixing and are also most strongly related to medium effects.  Because we are mostly concerned with the scaling of initial conditions and system size effects in this article, we limit ourselves to flow harmonics that develop primarily from linear response. However, we do explore the effectiveness of linear response across system size in Section \ref{sec:linear}. 

In order to determine the centrality binning (and the number of participants, Npart) we generated 3 million initial conditions for each collisional system and binned the initial total entropy.  Then, using these same initial conditions we were able to calculate multiparticle cumulants in Section \ref{sec:ec}. 

\subsection{System Size vs. Multiplicity}\label{sec:size}

Intuitively, a collision of two lead ions containing 208 nucleons each has a significantly larger system size than a collision of two oxygen ions containing 16 nucleons each.  However, it is not clear how well the number of participants in that collision, Npart, scales with the final multiplicity, M. Note that we use multiplicity here instead of $dN/dy$ because all results shown here are boost invariant so $dN/dy$ is somewhat misleading (although the results are comparable at mid-rapidity).  Additionally, multiplicity includes all particles even neutral ones. From the initial condition we can estimate the multiplicity using
\begin{equation}
M=\frac{S_0}{4}
\end{equation}
where $S_0$ is the total initial entropy.  From TRENTO one can obtain $S_0$ using 
\begin{equation}\label{eqn:mulfac}
S_0=a\cdot \tilde{s}
\end{equation}
 where $a$ is a free parameter that is tuned to obtain the correct multiplicity of all charged hardons in $0-5\%$ in PbPb collisions at $\sqrt{s_{NN}}=5.02$ TeV and $\tilde{s}$ is obtained from integrating over the entropy density of the initial condition itself.  In this paper we use $a=120\pm1$ for all collisions, assuming that $a$ is approximately independent of system size. We also note that viscosity plays a small role in the final multiplicity after running hydrodynamics, however, since we use the same viscosity in all runs, it is reasonable to make these comparisons directly from the initial conditions alone.

\begin{figure}[h]
\centering
\includegraphics[width=1\linewidth]{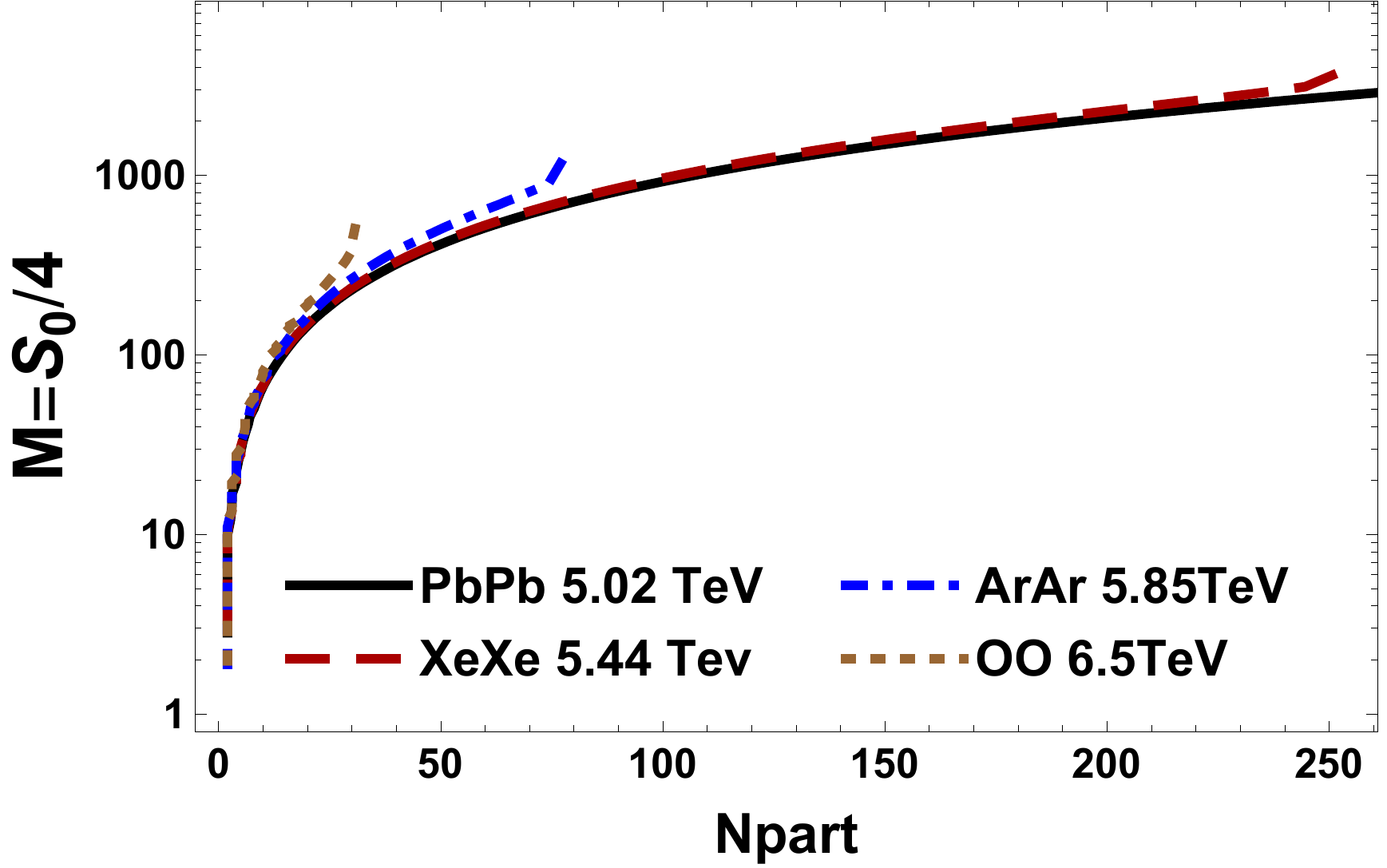}
\caption{(Color online) Estimate for the multiplicity, M, from the initial total entropy versus the number of participants, Npart.}
\label{fig:MvNpart}
\end{figure}

\begin{figure}[h]
	\centering
	\includegraphics[width=1\linewidth]{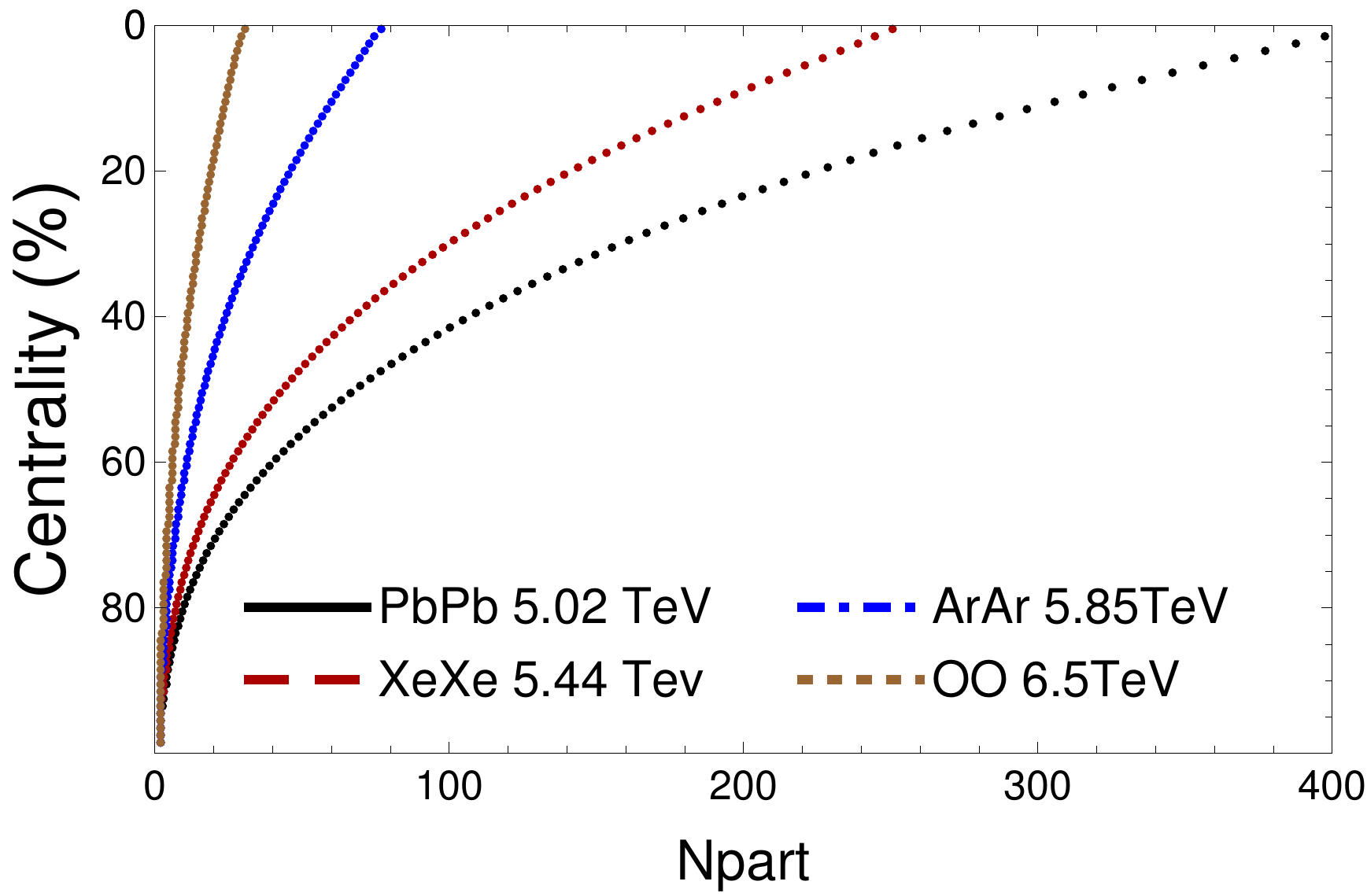}
	\caption{(Color online) Illustration of the centrality percentile (in 1\% bins) as a function of the number of participants, Npart, for different system sizes.}
	\label{fig:cenvNpart}
\end{figure}

In Fig.\ \ref{fig:MvNpart} the multiplicity vs. Npart is shown for PbPb, XeXe, ArAr, and OO collisions.   For central collisions there is generally a spike in multiplicity compared to Npart but otherwise there is generally a good scaling with Npart.  We note that this uptick in dN/dy has been seen already in experiments \cite{CMS:2018sks}. 
Thus, these results imply that Npart is useful but it has a slightly different scaling behavior than the multiplicity, especially in central collisions. Furthermore, we find that for the same Npart, smaller systems have a slightly larger multiplicity (especially in central collisions).  This feature is also reflected in the centrality distributions for the different system sizes shown in Fig.~\ref{fig:cenvNpart}.  Crucially, smaller systems at the same Npart correspond to much more central collisions than for larger systems.  In smaller systems, the same number of participants are distributed more compactly in a more central collision, while in larger systems those participants are distributed more dilutely across a wider, more peripheral interaction region.

While it is clear that OO collisions produce a smaller system than PbPb collisions, it does not mean that these small droplets of the Quark Gluon Plasma are identical even though they have the same final multiplicity. Here we calculate the radius of the initial condition using:
\begin{equation}
R^2=\frac{\int r^2 s(r,\phi) r drd\phi}{\int s(r,\phi) r drd\phi}
\end{equation}
where we note the ``radius" is not of a sphere but rather the variance from the center of mass. 
 
\begin{figure}[h]
\centering
\includegraphics[width=1\linewidth]{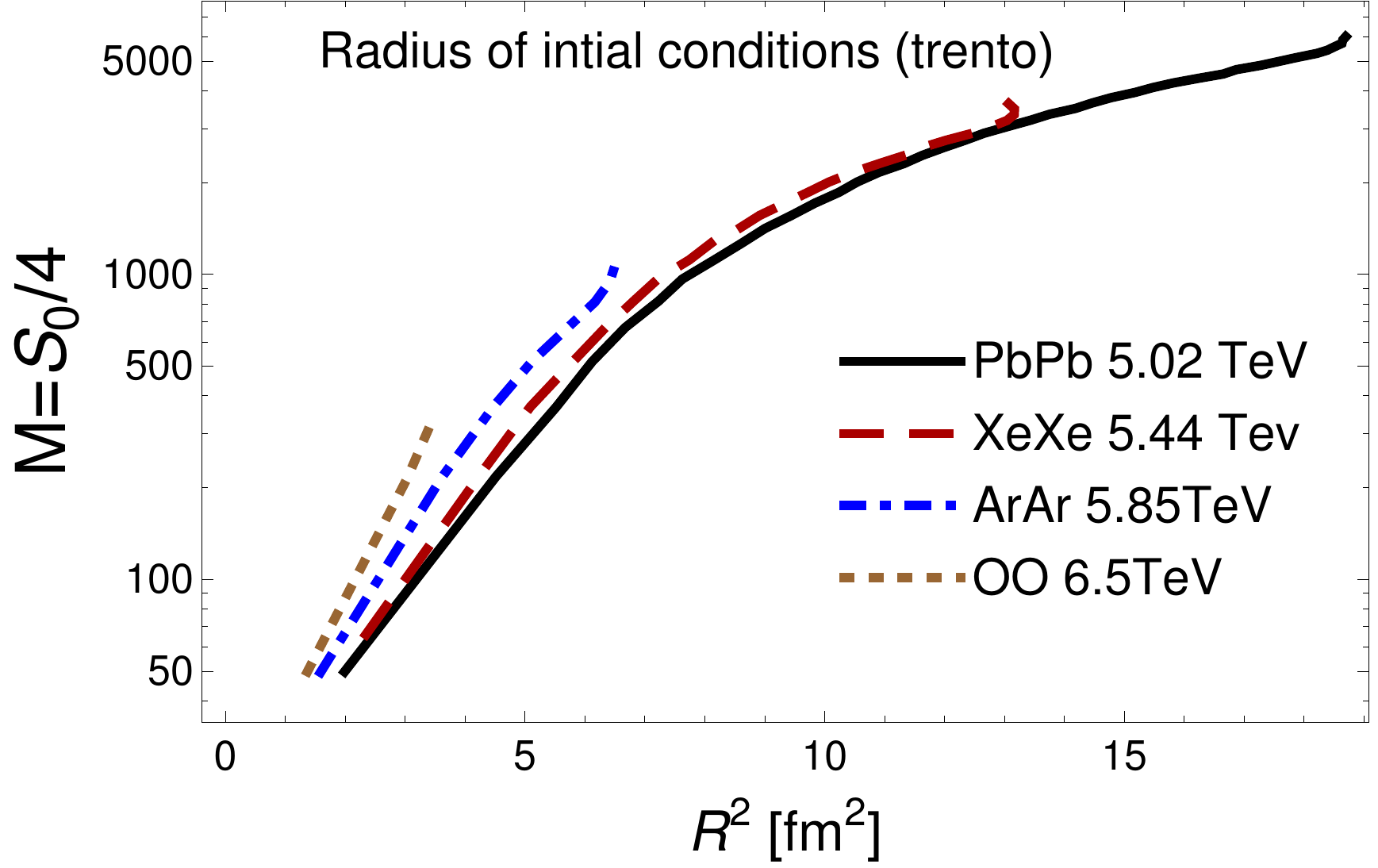}
\caption{(Color online) Estimate for the multiplicity, M, from the initial total entropy versus the radius of the initial condition.}
\label{fig:MvR}
\end{figure}
In Fig.\ \ref{fig:MvR} we plot the final multiplicity versus the initial R of the system.  For a fixed radius, since the multiplicity is directly related to the total entropy and that entropy is connected to the initial temperature, it is safe to say that systems with larger multiplicities reach higher temperatures on average.  Thus, we find that for the same system size, small systems reach much higher temperatures i.e. OO is a much hotter, tiny droplet of the QGP.  To estimate just how much hotter we can see that for R=3 fm, OO collisions have a $75\%$ larger initial entropy.  We find in this paper that these higher temperatures reached in small systems has a number of implications. 

\begin{figure}[h]
\centering
\includegraphics[width=1\linewidth]{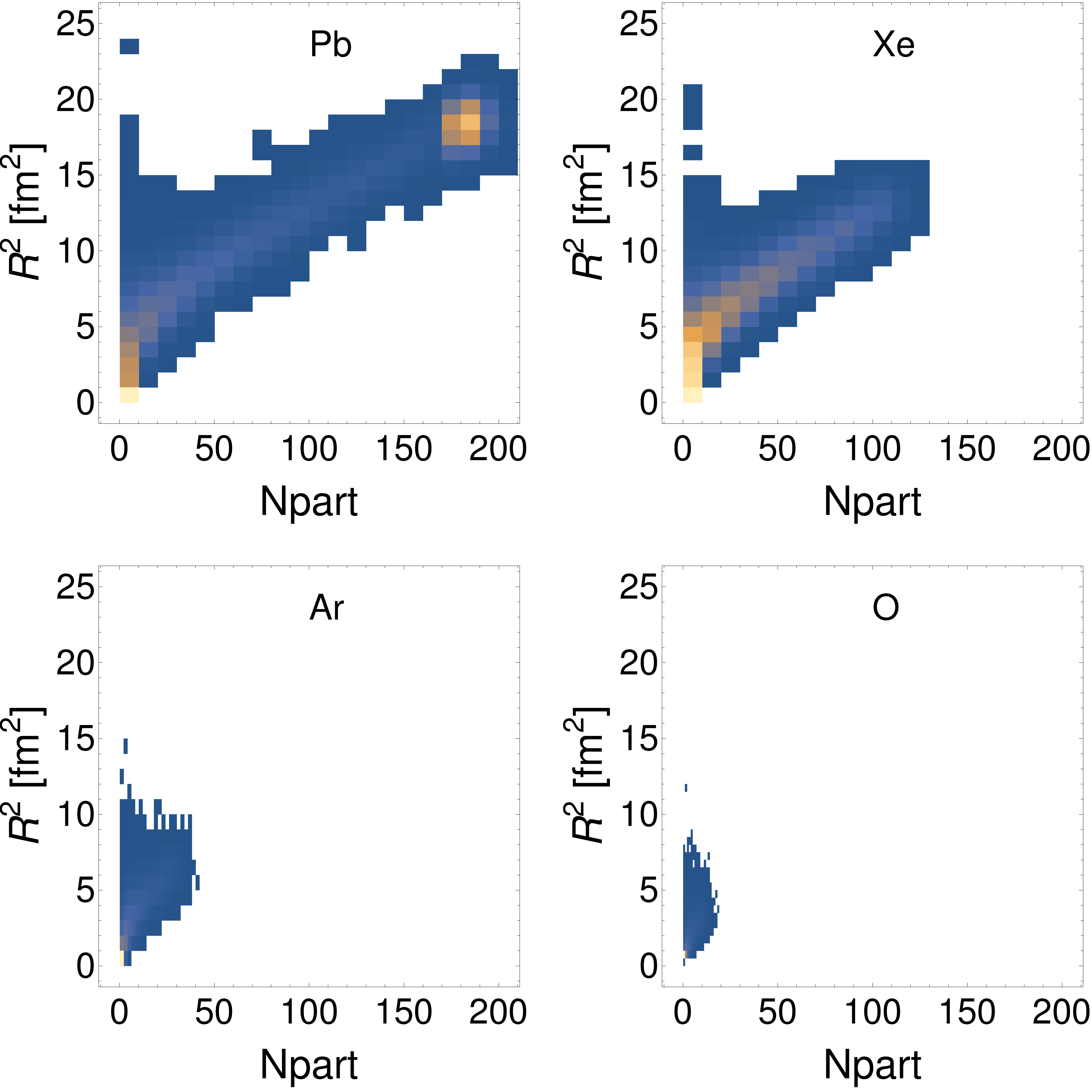}
\caption{(Color online) Density histograms of the radius versus the number of participants for PbPb, XeXe, ArAr, and OO collisions.}
\label{fig:radfluc}
\end{figure}
We do caution against generalizing that all PbPb collisions at a fixed Npart are larger in system size.  In Fig.\ \ref{fig:radfluc} we plot the density histogram of the radius vs. the number of participants.  At a fixed Npart,  while the radii in PbPb collisions  are generally larger than OO collisions, there are still large fluctuations in the radii of each collision on an event-by-event basis.  Furthermore, as one looks at the smallest multiplicities in all systems (e.g. $M\sim 50$), the variance in the radii between different collision types is the smallest (less than 1 fm difference).

\begin{figure}[h]
\centering
\includegraphics[width=1\linewidth]{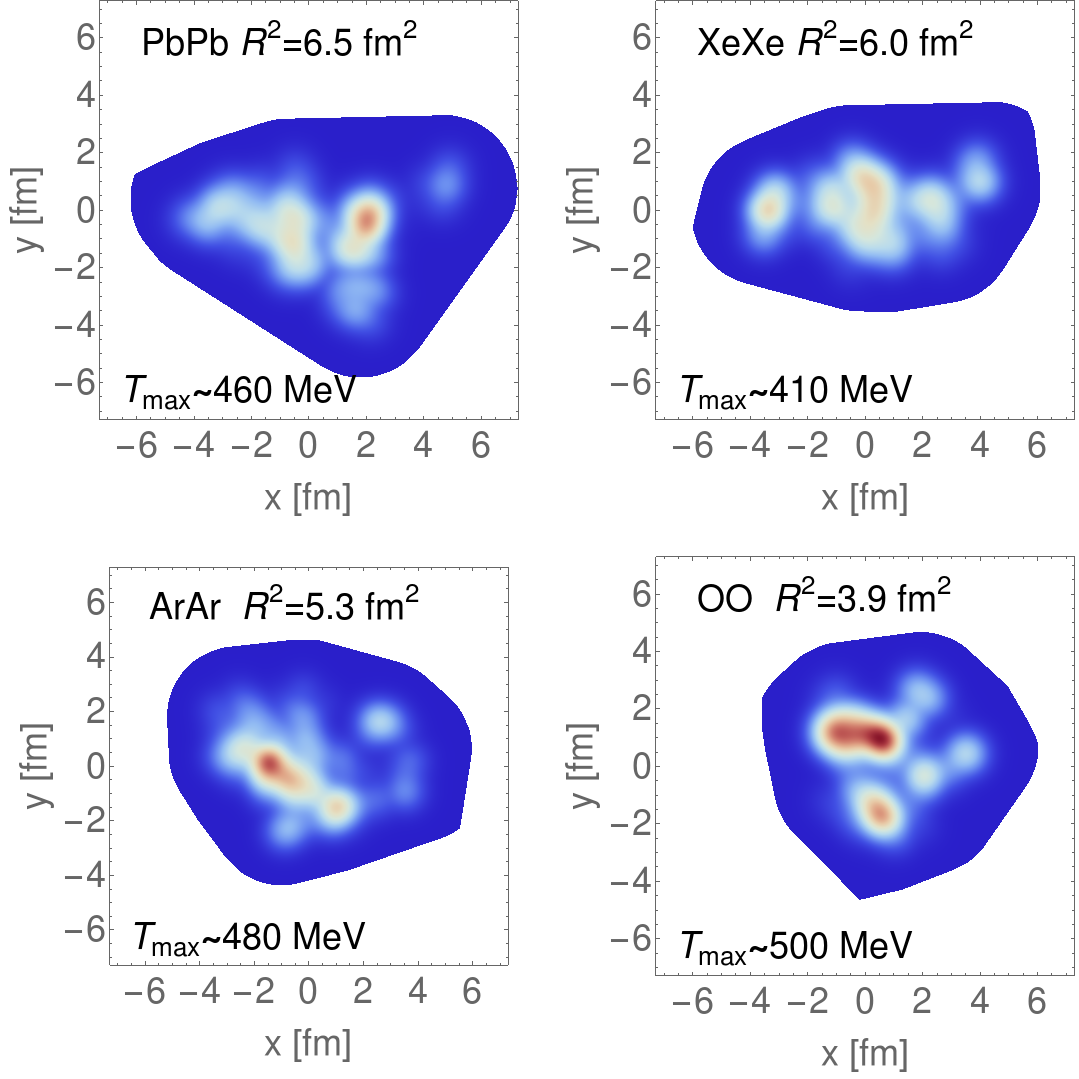}
\caption{(Color online) Entropy density plots of 4 example initial conditions with the initial entropy of $S_0=2225$ and radii that most closely reflect the average radius for their respective collision system. Red indicates the largest entropy densities and blue indicates less entropy density. All plots are scaled to the maximum entropy density (across all collision systems).}
\label{fig:pics}
\end{figure}

A good way to depict the differences between these systems is to select a fixed multiplicity (here we choose M=556, which is the event with the largest multiplicity we obtained from OO collisions) and compare a single event with that same multiplicity in the other collision types.  In order to avoid a statistically unlikely event, for PbPb, XeXe and ArAr collisions we calculate the average radius for events with  M=556 and then pick the event with the radius closest to the average.  The entropy density plot of these 4 events are shown in Fig.\ \ref{fig:pics}.  We can then correlation the largest hotspot in each event to the temperature using the equation of state and we find that the smallest system, OO, reaches a maximum temperature at its hotspot that is roughly $\Delta T\sim 50$ MeV larger than the PbPb collision.  

Our previous results that found that for the same multiplicity we're seeing a rounder impact region in OO collisions compared to 
PbPb can also be clearly visualized in Fig. \ref{fig:pics} as well.  The PbPb event has a clear elongated triangular shape, the XeXe event has a clear elliptical shape, the ArAr event has a smaller rounder shape, and the OO collision is an even smaller, rounder event. 

\section{Hydrodynamical Setup}\label{sec:hydro}

The event-by-event relativistic viscous hydrodynamic model is v-USPhydro \cite{Noronha-Hostler:2013gga,Noronha-Hostler:2014dqa} where the hydrodynamical setup is the same that was used for PbPb collisions at $\sqrt{s_{NN}}=5.02$ TeV \cite{Alba:2017hhe} and XeXe collisions at  $\sqrt{s_{NN}}=5.44$ TeV \cite{Giacalone:2017dud,Noronha-Hostler:2018zxc}. v-USPhydro is a Langrangian code based on Smooth Particle Hydrodynamics and passes well-known analytical test \cite{Marrochio:2013wla}.  The equation of state is the most up-to-date 2+1 Lattice QCD \cite{Borsanyi:2013bia} based equation of state, which is coupled to the most up-to-date particle data list, PDG16+ (as discussed in \cite{Alba:2017hhe,Alba:2017mqu}).  Only direct decays are considered using an adapted AZhydro code \cite{Kolb:2000sd,Kolb:2002ve,Kolb:2003dz}.  Here $\eta/s=0.047$, $\tau_0=0.6$ fm, $T_{FO}=150$ MeV, and $\zeta/s=0$.

In order to make direct comparisons with experimental data, cumulants of the flow harmonics \cite{Bilandzic:2010jr} are calculated using:
\begin{eqnarray}\label{eqn:cumulants}
\nonumber v_n\{2\}^2 &=& \langle v_n^2 \rangle, \\
\nonumber v_n\{4\}^4 &=& 2 \langle v_n^2\rangle^2 - \langle v_n^4\rangle, \\
\nonumber v_n\{6\}^6 &=& \frac{1}{4} \biggl [ \langle v_n^6 \rangle - 9 \langle v_n^2\rangle \langle v_n^4 \rangle + 12 \langle v_n^2\rangle^3 \biggr], \\
\nonumber v_n\{8\}^8 &=& \frac{1}{33} \biggl [ 144 \langle v_n^2\rangle^4 - 144 \langle v_n^2\rangle^2 \langle v_n^4 \rangle + 18 \langle v_n^4\rangle^2 \\
 &+& 16 \langle v_n^2 \rangle \langle v_n^6 \rangle - \langle v_n^8 \rangle \biggr ],\nonumber
\end{eqnarray}
where the moments of the $v_n$ distribution are used to calculate the cumulants. Here we use centrality rebinning as discussed in \cite{Gardim:2016nrr,Betz:2016ayq}.  We run 31,000 events for each different ion type and use jackknife resampling to determine statistical error. 

\section{Identified particles and $\langle p_T\rangle $}

In recent years  the mean transverse momentum, $\langle p_T\rangle$,  has garnered a significant amount of attention because it was found that IP-Glasma initial conditions generally produce too much $\langle p_T\rangle$.  However, the inclusion of a larger peaked bulk viscosity close to the phase transition appeared to resolve the issue \cite{Ryu:2015vwa}.  That being said, TRENTO initial conditions did not appear to need such a larger bulk viscosity as discussed in \cite{Bernhard:2016tnd} nor does averaged (smooth) Glauber initial conditions+anistotropic hydrodynamics \cite{Alqahtani:2017tnq}. Thus, studying the scaling of $\langle p_T\rangle$ across system sizes may provide further insight into transport coefficients.

The difference between PbPb and XeXe collisions were quite small and the previous hydrodynamical setup with only the inclusion of a fixed $\eta/s$ \cite{Giacalone:2017dud,Noronha-Hostler:2018zxc} was able to predict the experimental data accurately \cite{Acharya:2018eaq}.  However, a comparison to an even smaller system size may provide new knowledge about the need (or lack thereof) of bulk viscosity.  Thus, here we provide baseline calculations that only include the same fixed $\eta/s$ as in \cite{Giacalone:2017dud,Noronha-Hostler:2018zxc}.  Deviations from our predictions may prove interesting to study bulk viscosity further. 

\begin{figure*}[h]
\centering
\begin{tabular}{c c}
\includegraphics[width=0.5\linewidth]{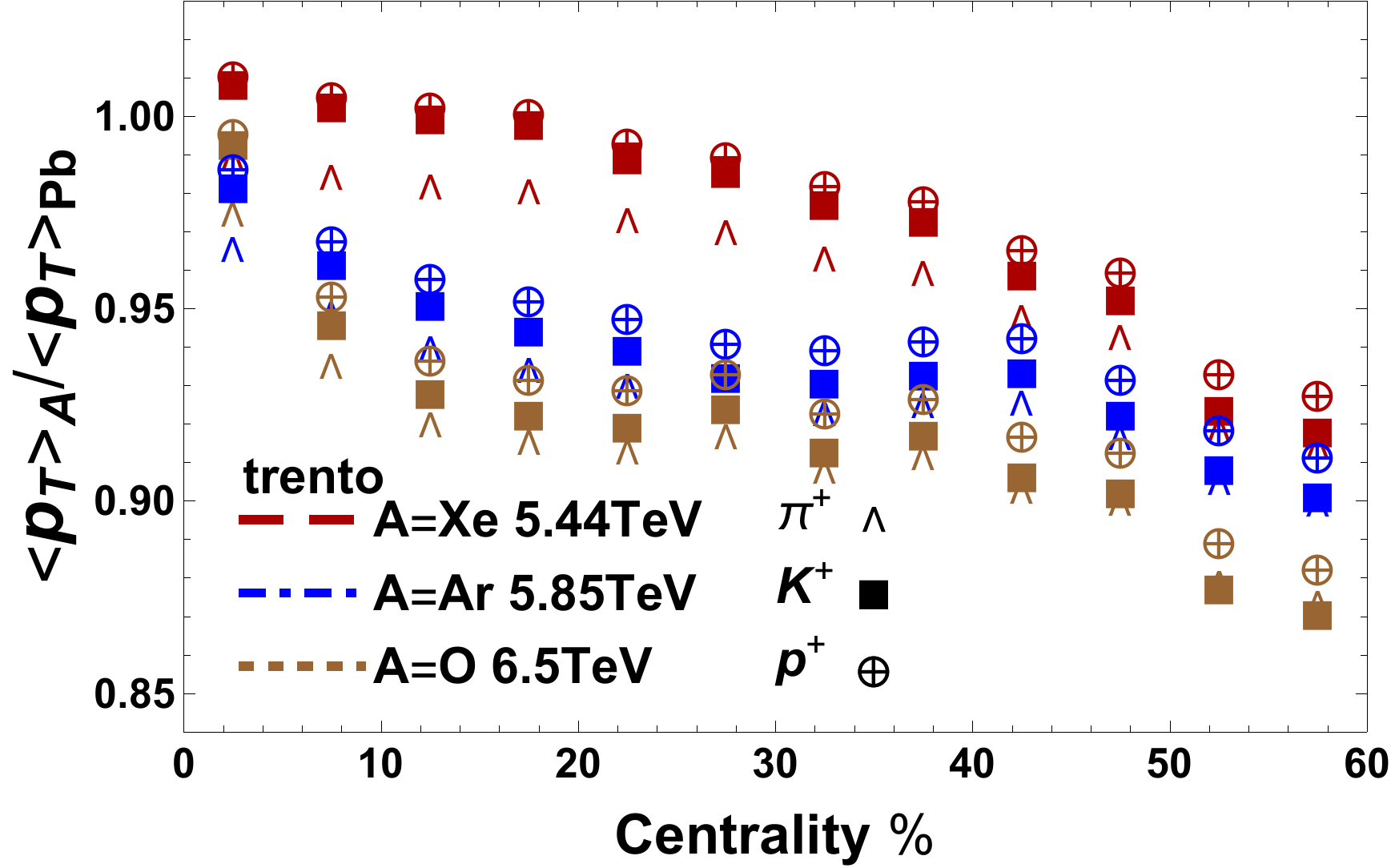} & \includegraphics[width=0.5\linewidth]{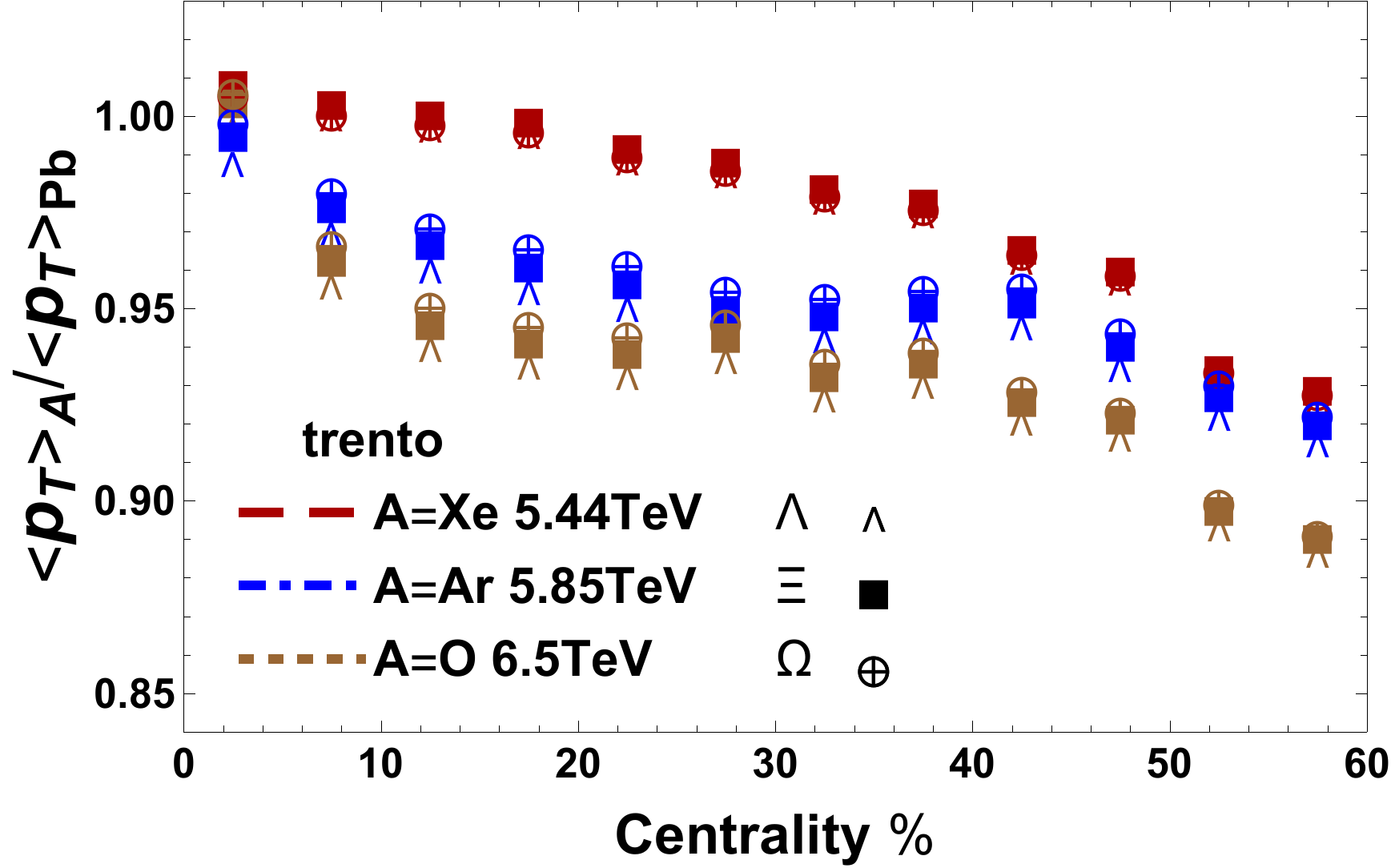}
\end{tabular}
\caption{(Color online) Ratio of $\langle p_T\rangle$ of AA collisions to $\langle p_T\rangle$ in PbPb collisions across centrality.   }
\label{fig:meanpt}
\end{figure*}

In Fig.\ \ref{fig:meanpt} we find that $\langle p_T\rangle$ should decrease with decreasing system size and that more peripheral collisions see the largest decrease.  The ratio of $\langle p_T\rangle_A/\langle p_T\rangle_{Pb}$ does not appear to be strongly dependent on the type of identified particles either.  Certainly, the system size plays the largest role. 

\begin{figure}[h]
\centering
\includegraphics[width=\linewidth]{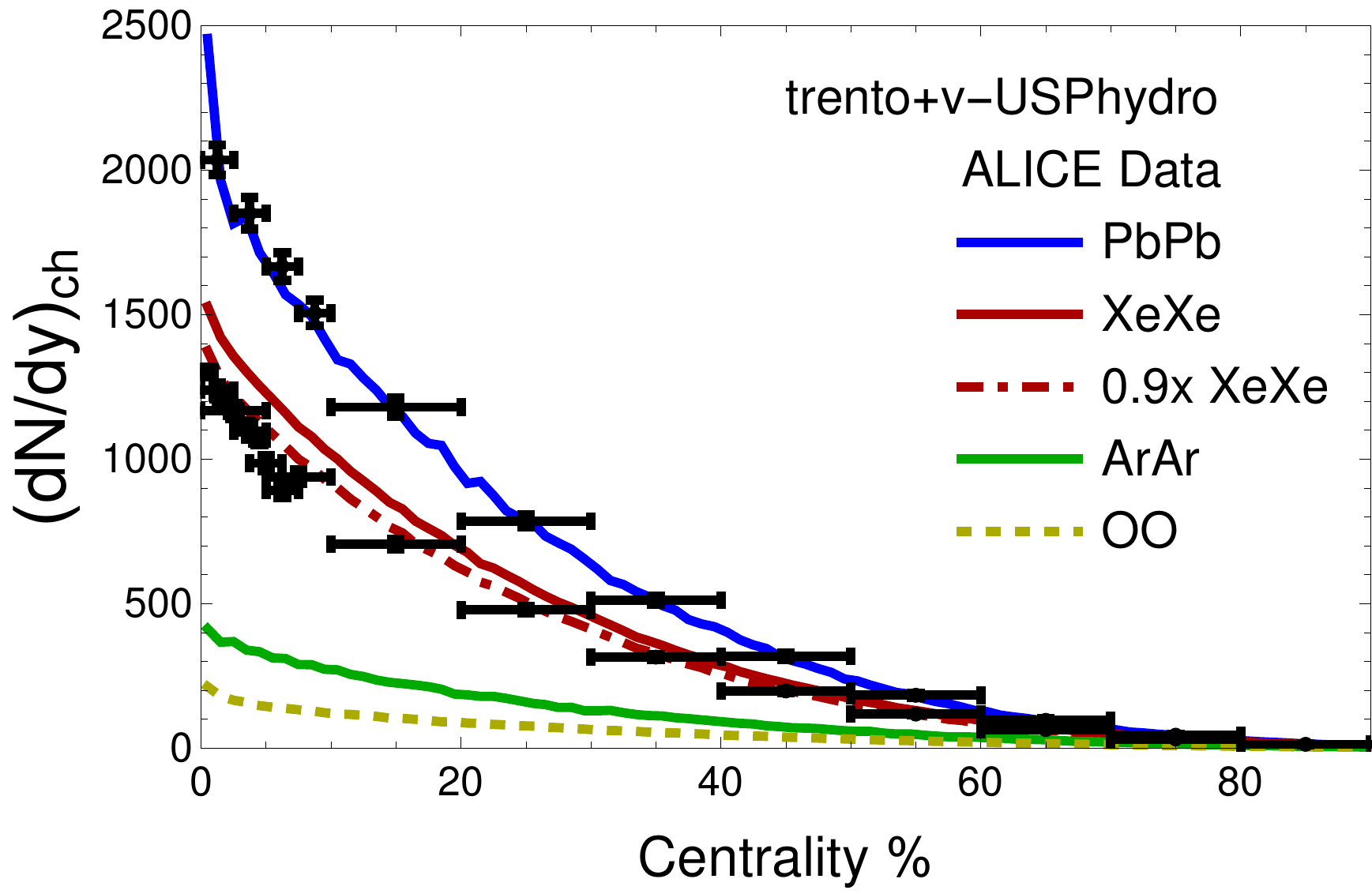} 
\caption{(Color online) Comparison of $\left(dN/dy\right)_{chg}$ across centrality for PbPb, XeXe, ArAr, and OO collisions from trento+v-USPhydro.  ALICE experimental data from PbPb \cite{Adam:2015ptt} and XeXe \cite{Acharya:2018hhy} is shown.   }
\label{fig:dndy_chg}
\end{figure}

We also compare the number of charged particles i.e. $\left(dN/dy\right)_{ch}$ to experimental data at ALICE for PbPb and XeXe collisions (while making predictions for ArAr and OO collisions).  We note here that in the original TRENTO+v-USPhydro XeXe paper \cite{Giacalone:2017dud} the assumption was made that the same multiplying constant $a$ in Eq.\ (\ref{eqn:mulfac}) could be used for XeXe as was used previously for PbPb.  Correspondingly, in Eq.\ (\ref{eqn:mulfac}) we have used the same constant $a$ for ArAr and OO, independent of system size.  As mentioned previously, this constant, $a$, is a free parameter that is tuned to experimental data normally in the most central collisions.  In Fig.\ \ref{fig:dndy_chg} we compare $\left(dN/dy\right)_{ch}$ for our predictions to ALICE data from PbPb \cite{Adam:2015ptt} and XeXe \cite{Acharya:2018hhy}.  Surprisingly, we find that our prediction for XeXe $\left(dN/dy\right)_{ch}$ roughly overshoots the data but rescaling the multiplicity down by a factor of 0.9 can then reproduce the ALICE data.  We then reran the hydrodynamic events with $a=108$, which was guessed from multiplying our original constant by 0.9 and from there we can easily reproduce the ALICE data (as shown).  We also confirmed that we did not see a significant change in our flow results from this small change. We do, however, note that we are unsure of what this implies for even smaller systems of ArAr and OO nor do we have a strong physics motivation for why a smaller system should require a smaller multiplying constant. We hope that future runs in small systems could shed more light on this issue.

\section{Linear Response}\label{sec:linear}

From the initial conditions alone we can learn quite a bit about the scaling behavior that one is to expect for the different types of ion collisions.  It has been shown that in large system sizes of AuAu and PbPb collisions that a strong linear mapping exists on an event-by-event basis between the initial elliptical and triangular eccentricities and the final $v_2$ and $v_3$, respectively \cite{Teaney:2010vd,Gardim:2011xv,Niemi:2012aj,Teaney:2012ke,Qiu:2011iv,Gardim:2014tya,Betz:2016ayq}. Furthermore, the final flow harmonics appear to be sensitive to only the large scale, geometrical structure, not the small scale fluctuations \cite{Gardim:2017ruc,Noronha-Hostler:2015coa,Mazeliauskas:2015vea,Konchakovski:2014fya}. Thus, focusing on how the eccentricities on an event-by-event basis scale with either the centrality or Npart can provide hints of what one would expect the flow harmonics to experience as well (although the final flow harmonics enfold all the non-linearities that are also exhibited within hydrodynamics so certain deviations are expected).  In fact, one question we will probe in this paper is how these non-linearities scale with the system size but to do that, we first need to understand the eccentricities. 
 
\begin{figure*}[h]
\centering
\includegraphics[width=1\linewidth]{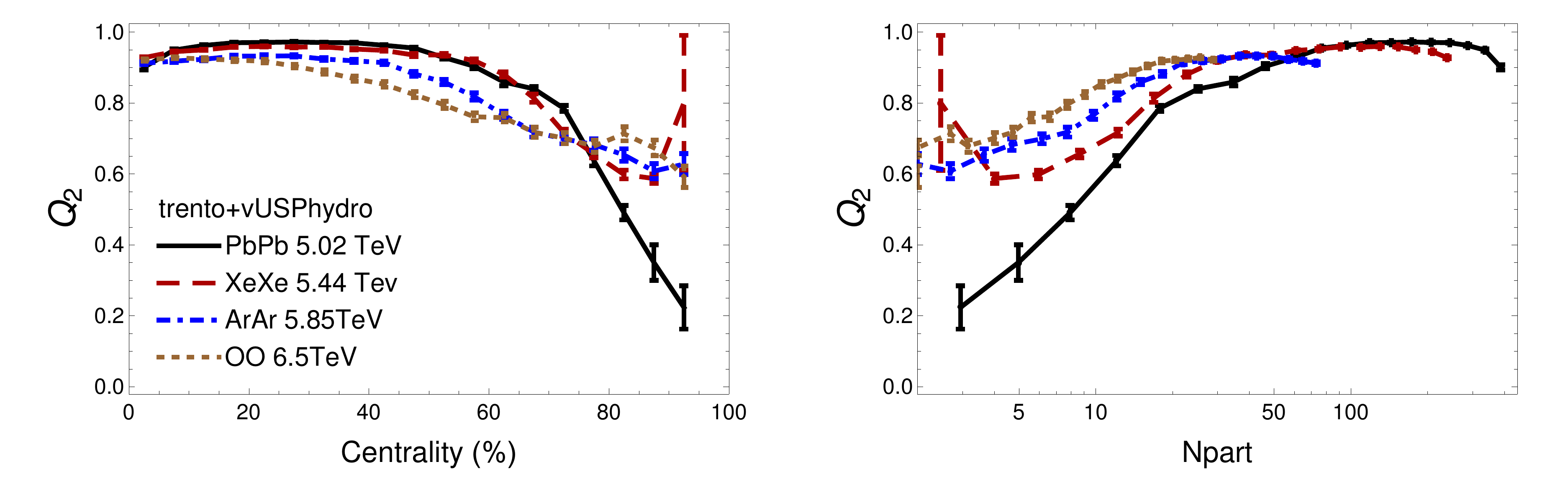}
\caption{(Color online) Pearson coefficient as defined in Eq. (\ref{eqn:pear}) for the elliptical flow for the four different collisional systems scaled by the centrality (left) and Npart (right).}
\label{fig:Q2}
\end{figure*}

\begin{figure*}[h]
\centering
\includegraphics[width=1\linewidth]{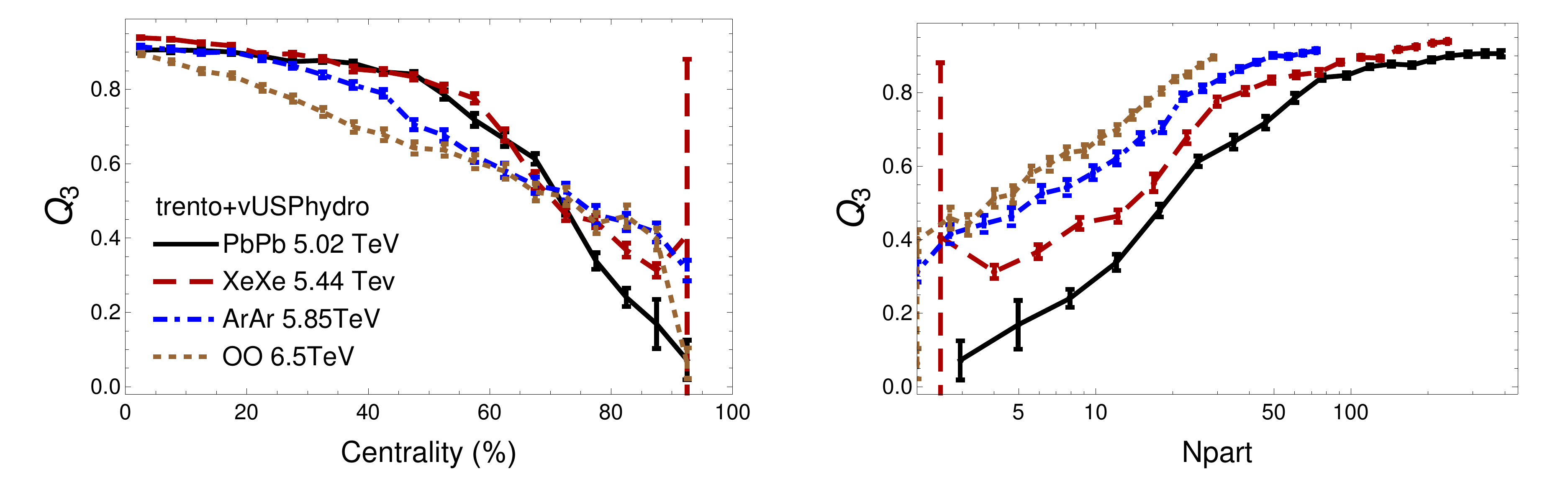}
\caption{(Color online) Pearson coefficient as defined in Eq. (\ref{eqn:pear}) for the triangular flow for the four different collisional systems scaled by the centrality (left) and Npart (right).}
\label{fig:Q3}
\end{figure*}

\subsection{Mapping}

As discussed previously, there is generally a dominate linear mapping between the initial conditions quantified by eccentricities and the final flow harmonics quantified by flow harmonics. Thus, for linear response one could predict the final flow harmonic via
\begin{equation}\label{eqn:linear}
V_n=\gamma_n \mathcal{E}_n
\end{equation}
where $\gamma_n$ is the coefficient in the purely linear case, which can be obtained using:
\begin{equation}\label{eqn:gamma1}
\gamma_n=\frac{Re\left( \langle V_n\mathcal{E}_n^*\rangle\right)}{\langle |\varepsilon_n|^2\rangle}.
\end{equation}
Once $\gamma_n$ is obtained, then observables that are only sensitive to linear response could be predicted from eccentricities alone.  Unlike in  \cite{Noronha-Hostler:2015dbi} where the linear coefficient is referred to as $\kappa_n$, we use $\gamma_n$ here to distinguish from the non-linear coefficients discussed in Sec.\ \ref{sec:cubic}.

To quantify the strength of this linear mapping depending on system size, we use a Pearson coefficient that compares the the flow vectors $\left\{v_n,\psi_n\right\}$ with the eccentricities $\left\{\varepsilon_n,\phi_n\right\}$ \cite{Gardim:2011xv,Gardim:2014tya,Betz:2016ayq}.  Note that this takes into account both the magnitude and the angle of each event (and harmonic):
\begin{equation}\label{eqn:pear}
Q_n=\frac{\langle v_n \varepsilon_n \cos\left(n\left[\psi_n-\phi_n\right]\right)\rangle}{\sqrt{\langle |\varepsilon_n|^2\rangle \langle  |v_n|^2 \rangle}}.
\end{equation}
As $Q_n\rightarrow 1$, a perfect linear mapping between the eccentricity and the final flow harmonic exists whereas if $Q_n\rightarrow 0$ other contributions beyond linear response are contributing to the final flow harmonic.  We note that it has already been shown that for peripheral PbPb collisions nonlinear response (in the form of cubic response) is necessary to understand the elliptical fluctuations data \cite{Noronha-Hostler:2015dbi}.  We also remark that a Pearson coefficient can be used to quantify any predictor for the final flow harmonics, even non-linear mapping, which has been previously explored in a number of papers \cite{Gardim:2011xv,Gardim:2014tya,Noronha-Hostler:2015dbi}.

In Fig.\ \ref{fig:Q2}-\ref{fig:Q3} the Pearson coefficients for the linear mapping from $\varepsilon_2\rightarrow v_2$  and $\varepsilon_3\rightarrow v_3$ are shown scaled by the centrality binning (left) and Npart (right).  A clear pattern emerges for both the elliptical and triangular flows: the linear response works best for central to mid-central collisions (when scaled by the centrality) but in the centrality range from $\sim 30-70\%$ there is a clear hierarchy in which the largest systems have the cleanest linear mapping from the eccentricities to the final state.  For peripheral collisions beyond $>75\%$ this hierarchy flips and the smallest system has the best linear mapping.  

In order to understand this behavior, one can look at the Npart scaling and we find that a consistent hierarchy arises.  The smallest system actually has the best linear mapping when scaled by Npart whereas the largest system would need some sort of nonlinear effects in order to predict the final flow data. This aligns with previous results in \cite{Noronha-Hostler:2015dbi} that found the necessity of cubic response in peripheral collisions.  For instance, if we compare an Npart=10, this occurs at $\sim 80\%$ centrality for PbPb collisions, and $42-45\%$ centrality for OO collisions.

\subsection{Predictions from linear scaling}\label{sec:ec}

\begin{figure*}[h]
\centering
\includegraphics[width=\linewidth]{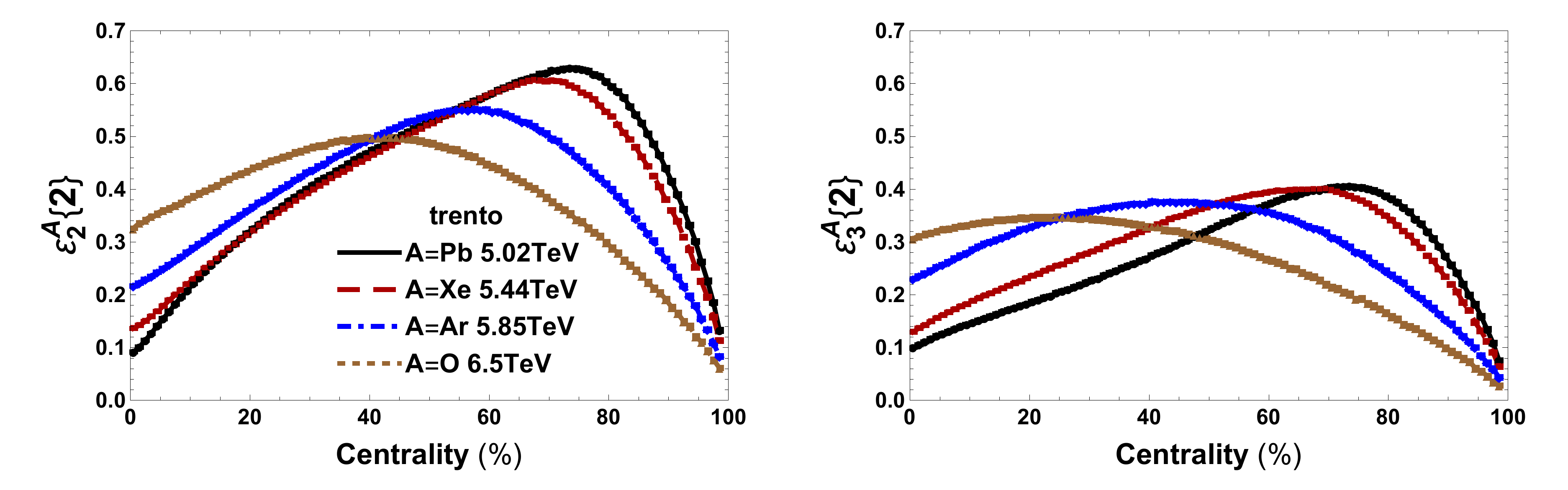}\\
\includegraphics[width=\linewidth]{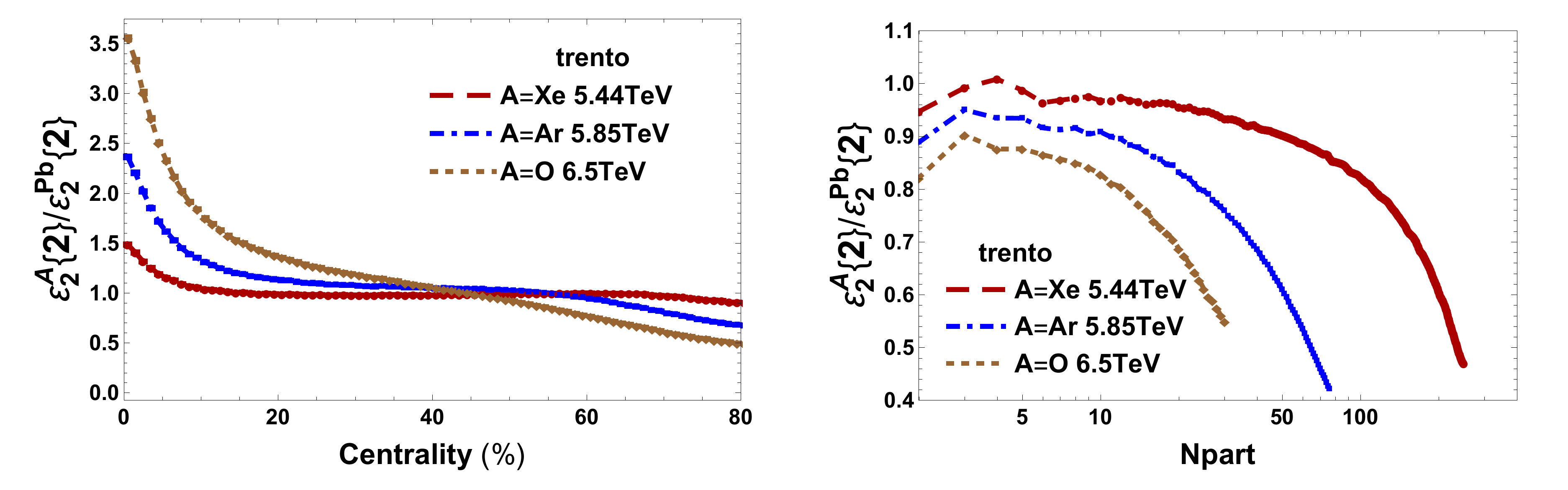}\\
\includegraphics[width=\linewidth]{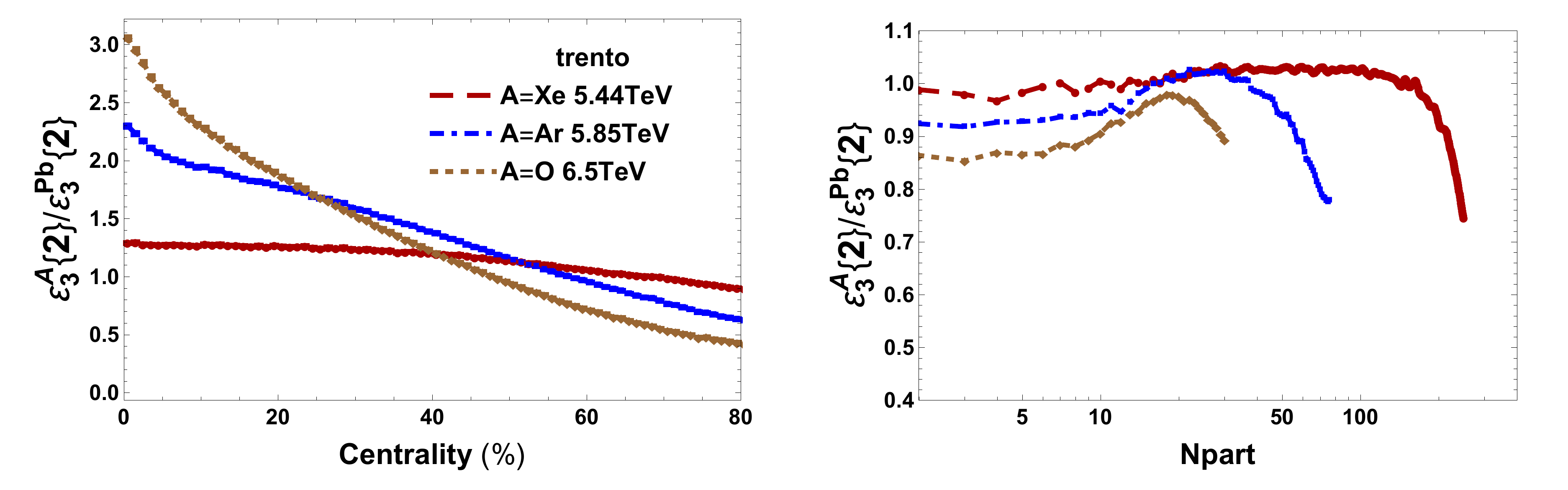}
 \caption{(Color online) Absolute values of $\varepsilon_2\{2\}$ and $\varepsilon_3\{2\}$ across centrality (top) for all different collision systems. Ratio of the eccentricities $\varepsilon_2$ (middle) and  $\varepsilon_3$ (bottom) where smaller ions are divided by the results in PbPb collisions.  These are scaled by the centrality (left) and Npart (right) }
\label{fig:eccrat}
\end{figure*}

Before we discuss the scaling of nonlinear response with system size, we discuss the results one would expect if only linear response was valid.  First we plot just the absolute values of $\varepsilon_2\{2\}$ and $\varepsilon_3\{2\}$ across centrality in the top of Fig.\ \ref{fig:eccrat}.  A few noticeable take aways are that:
\begin{itemize}
\item  $\varepsilon_3\{2\}$ is flatter across centrality than $\varepsilon_2\{2\}$ regardless of system size
\item Both  $\varepsilon_2\{2\}$ and $\varepsilon_3\{2\}$ display a peak in their values at mid-centrality but as the system size is decreased that peak shifts to more peripheral collisions.  Generally, the peak shifts less in $\varepsilon_2\{2\}$ than in $\varepsilon_3\{2\}$ .
\item Regardless of the system size, the peak in $\varepsilon_2\{2\}$ is steeper than  $\varepsilon_3\{2\}$ (again related to the smaller centrality dependence for $\varepsilon_3\{2\}$).  Additionally, the peak in $\varepsilon_2\{2\}$ occurs at more central collisions than the corresponding peak in $\varepsilon_3\{2\}$. 
\item As the system size decreases it appears that both eccentricities are less sensitive to centrality dependence.  
\end{itemize}

Because the $\gamma_n$ coefficient  shown in Eq.\ (\ref{eqn:linear}) is dependent on the initial conditions and medium properties, we prefer to plot ratios of eccentricities such that their $\gamma_n$'s should cancel out.  We do note, however, that $\gamma_n$'s are different for each flow harmonic so ratios of eccentricities of different order in n do not cancel exactly. Furthermore, we have already noted in the previous section that the linear response changes as system size is decreased, which implies that the $\gamma_n$'s are also system size dependent so we do not expect exact scaling across system size as well.  Instead these results can be used a baseline to compare how different the expectations of the eccentricities are versus the final flow harmonics.

With these caveats in mind, in Fig.\ \ref{fig:eccrat} the ratio of the elliptical eccentricities are shown versus centrality (left) and Npart (right) divided by the largest system, PbPb collisions. 
Immediately one sees quite different trends depending if one plots versus Centrality or Npart.  Plotting versus centrality, a crossing is exhibited at $\sim45\%$ centrality where for very central collisions the smaller systems have the largest $\varepsilon_n$'s. In contrast,  in peripheral collisions the smallest systems have the smallest eccentricities.  Thus, if linear response is valid, one would expect a similar trend in the final flow harmonics.  Then plotting versus Npart, a different story emerges where the smaller the system, the smaller the $\varepsilon_n$'s. In Section \ref{sec:results} we find that the final flow harmonics exhibit quite a different behavior due to non-linear response. Both elliptical and triangular eccentricities exhibit similar behaviors across centrality/Npart but we do find that the triangular eccentricities are generally less sensitive to system size changes. When plotting versus Npart the general conclusion can be drawn that larger systems generate the largest $\varepsilon_n$'s and that this is most evident for the largest Npart.

\begin{figure*}[h]
\centering
\begin{tabular}{c c}
\includegraphics[width=0.5\linewidth]{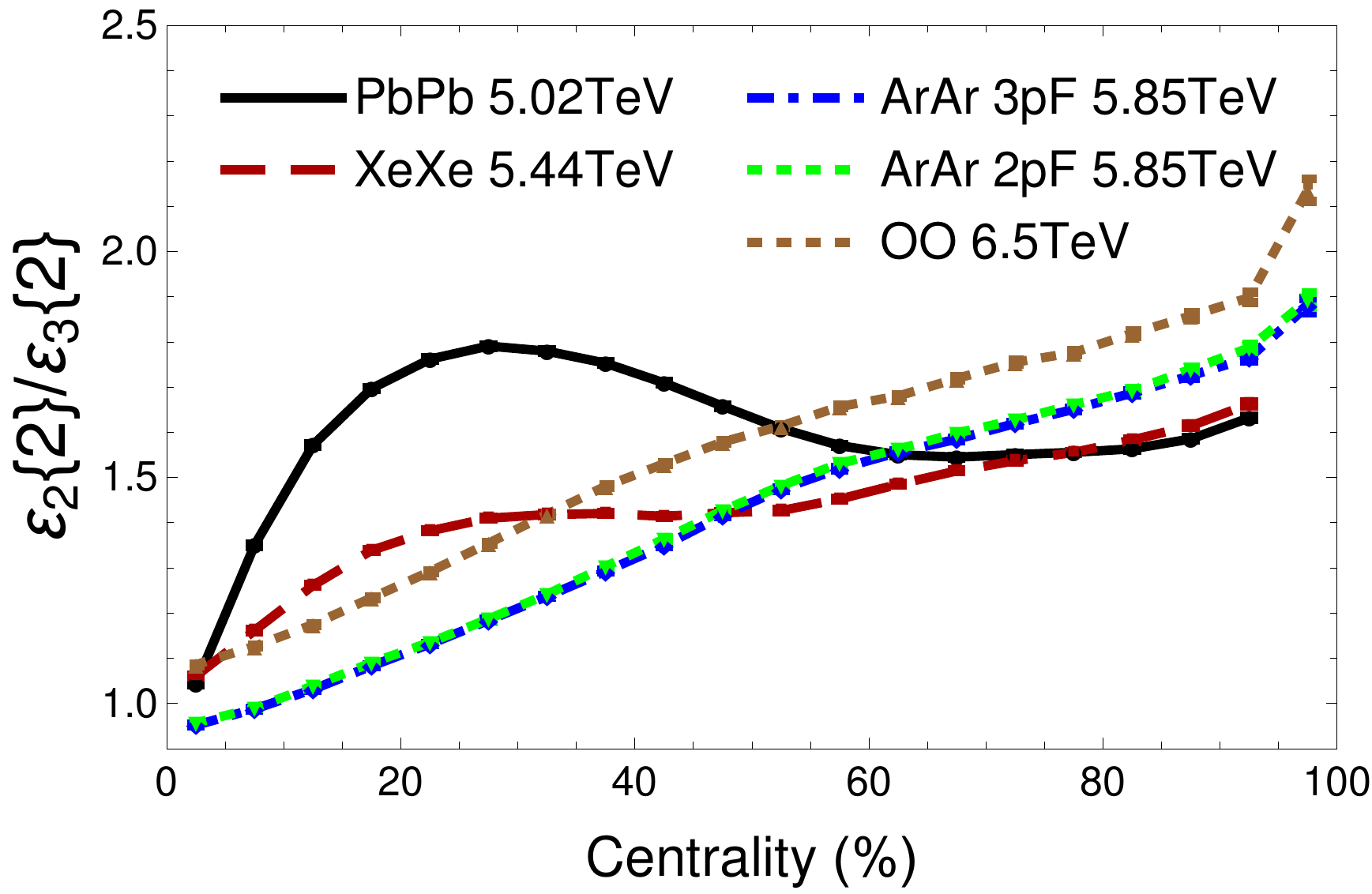} & \includegraphics[width=0.5\linewidth]{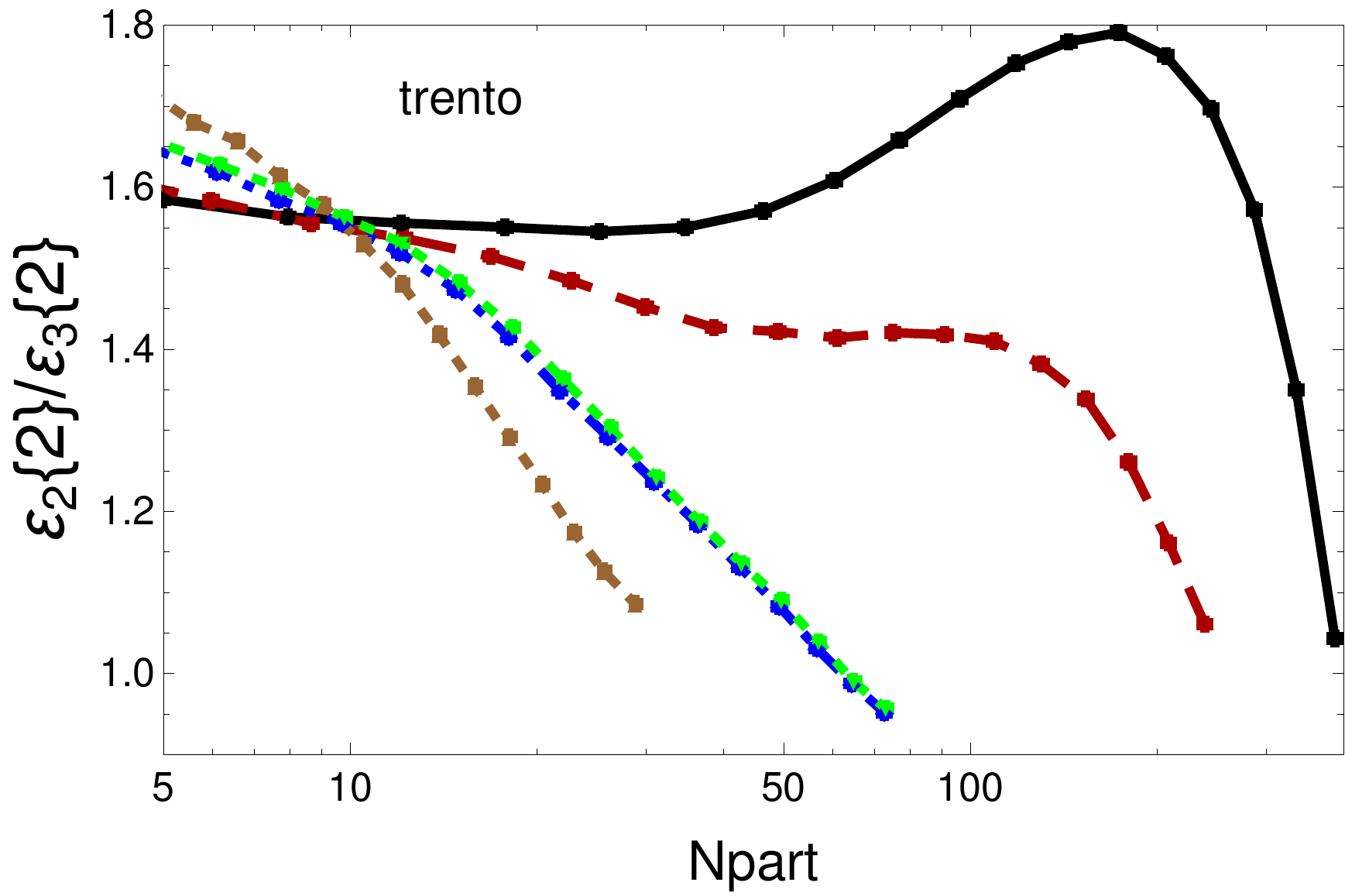}
\end{tabular}
\caption{(Color online) Ratio of $\varepsilon_2\left\{2\right\}/\varepsilon_3\left\{2\right\}$  across centrality (left) and compared to Npart (right) for the four collision types.  The abbreviation 3pF signifies the 3 parameter Fermi distribution fit for the Wood-Saxon distribution in Eq. (\ref{e:3PF}) and 2pF signifies the 2 parameter Fermi distribution fit in Eq. (\ref{e:2PF}). }
\label{fig:v2v3}
\end{figure*}

While we have already noted that the $\gamma_n$'s do not cancel for ratios of mixed order, we do still explore the  $\varepsilon_2\left\{2\right\}$ to $\varepsilon_3\left\{2\right\}$  relationship because it can be useful to constrain initial conditions \cite{Retinskaya:2013gca}.  
In Fig.\ \ref{fig:v2v3} we plot the ratio of $\varepsilon_2\left\{2\right\}/\varepsilon_3\left\{2\right\}$ and find a somewhat surprising result.  For PbPb and XeXe there is very similar behavior across both centrality and in Npart a peak is displayed in $\varepsilon_2\left\{2\right\}/\varepsilon_3\left\{2\right\}$ where the same peak is seen in the absolute value of $\varepsilon_2\left\{2\right\}$ in Fig.\ \ref{fig:eccrat}.  Because the dependence of eccentricities on centrality weakens as one goes to smaller systems, ArAr and OO collisions no longer has a clear peak across centrality for $\varepsilon_2\left\{2\right\}/\varepsilon_3\left\{2\right\}$.  Additionally, when plotted versus centrality $\varepsilon_2\left\{2\right\}/\varepsilon_3\left\{2\right\}$ is larger for OO collisions compared to ArAr collisions, however, we conclude this is simply because for the same centrality, OO collisions have a smaller Npart (and Npart is inversely proportional to $\varepsilon_2\left\{2\right\}/\varepsilon_3\left\{2\right\}$).
One final point of interest is that at Npart$\sim 10$ the ratio of $\varepsilon_2\left\{2\right\}/\varepsilon_3\left\{2\right\}$ is equal for all systems.

In \cite{Giacalone:2017uqx} it was found that the ratios of multi-particle cumulants can be used as an important constraint of initial conditions in heavy ion collisions.  Specifically, it appears that Glauber conditions systematically underpredicted all the experimental data and most initial conditions underpredict $v_3\{4\}/v_3\{2\}$ in central collisions.  Thus, studying specifically $v_n\{4\}/v_n\{2\}$, which is a proxy for the amount of fluctuations in the system, is vital in understanding the predictive power of the known initial conditions on the market. Furthermore, the ratio $v_n\{4\}/v_n\{2\}$ is primarily generated from linear response (although we will discuss those caveats in Sec. \ref{sec:cubic} ) so the ratio $\varepsilon_n\{4\}/\varepsilon_n\{2\}$ is a uniquely powerful observable that provide direct information about the initial conditions themselves independently of information about the medium properties. Finally, these observables  require a large amount of statistics, which require long run times in relativistic hydrodynamics, thus, it can be often easier to compare the eccentricities directly.
\begin{figure*}[h]
\centering
\begin{tabular}{c c}
\includegraphics[width=0.5\linewidth]{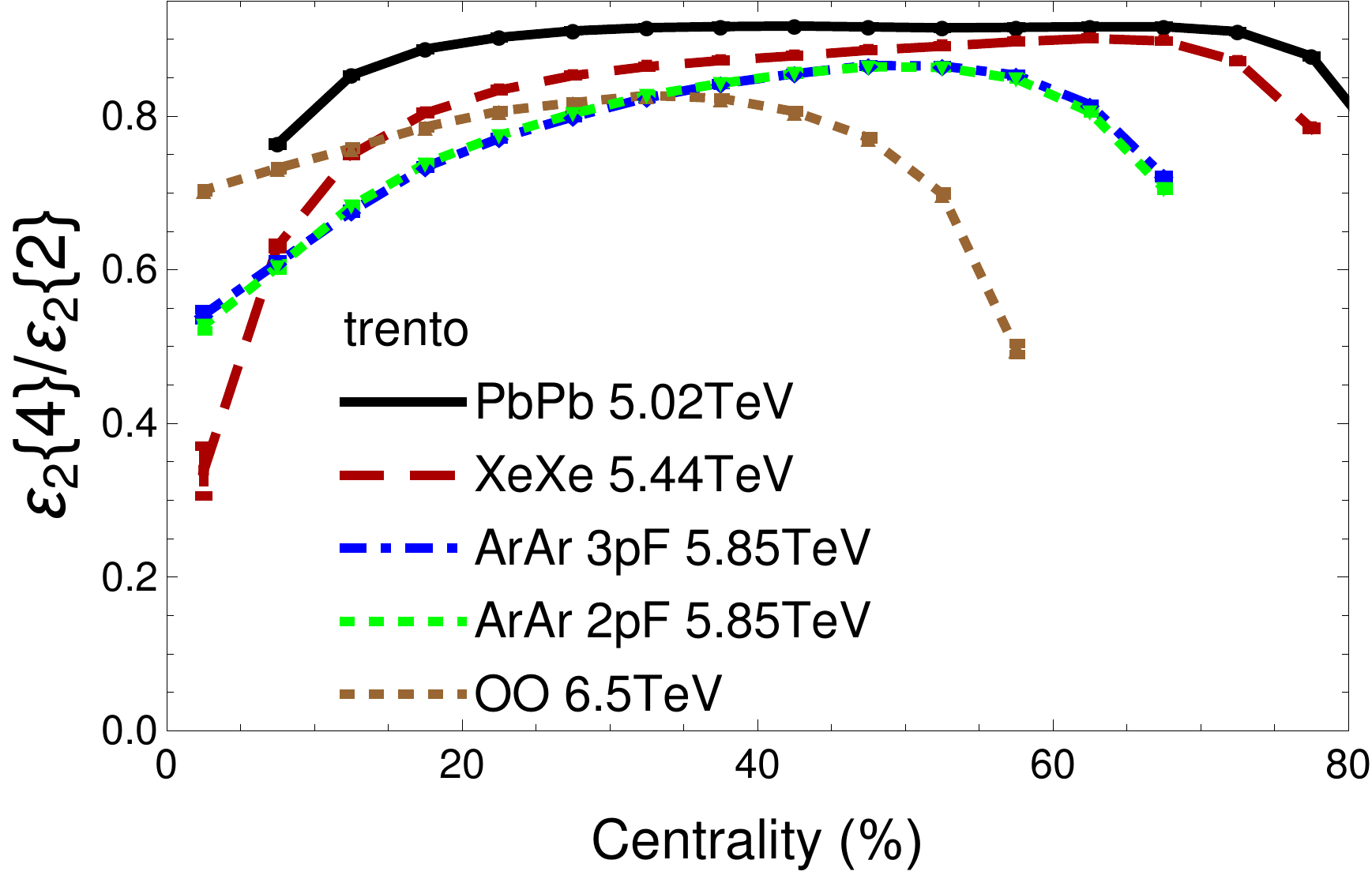} & \includegraphics[width=0.5\linewidth]{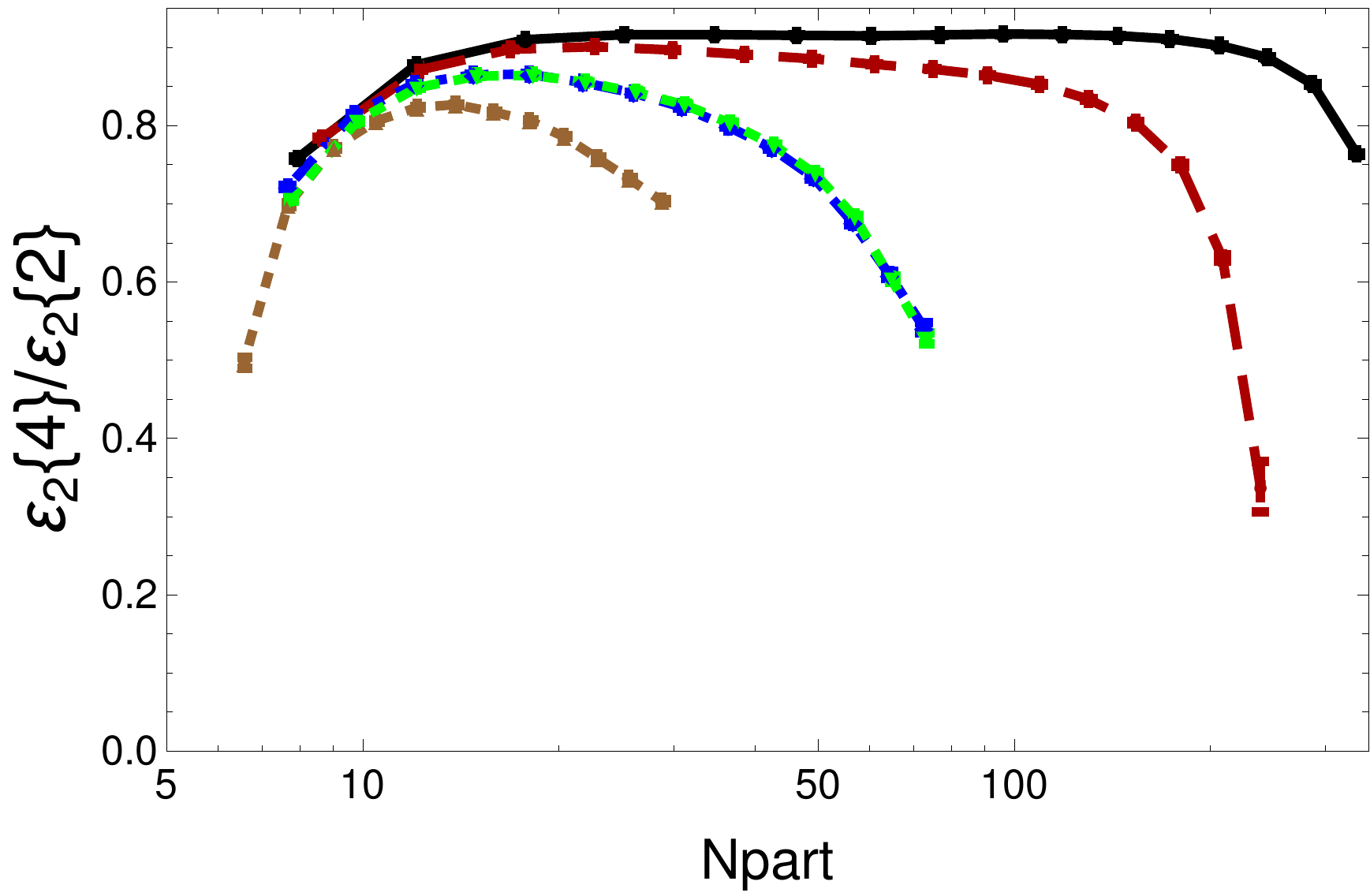}
\end{tabular}
\caption{(Color online) Ratio of $\varepsilon_2\{4\}/\varepsilon_2\{4\}$ for all system sizes plotted versus centrality (left) and Npart (right).  Note that 3pF denotes a 3 parameter Wood Saxon fit and 2pF denotes the typical two parameter Wood Saxon. }
\label{fig:cumu_v2}
\end{figure*}

In Fig.\ \ref{fig:cumu_v2} the ratio $\varepsilon_n\{4\}/\varepsilon_n\{2\}$ (see Eq.\ (\ref{eqn:cumulants}), which can be calculated in the same manner using the eccentricities instead of the flow harmonics ) is shown plotted versus centrality and Npart.  The comparison versus centrality is rather difficult to draw direct conclusions from but we do remark that generally there are more fluctuations in smaller systems (note that as $\varepsilon_n\{4\}/\varepsilon_n\{2\}\rightarrow 1$ there are less fluctuations in the system i.e. one would expect a narrower distribution around the mean of $v_2$ and a smaller value of $\varepsilon_n\{4\}/\varepsilon_n\{2\}$ implies a wider distribution around the mean). The comparison of $\varepsilon_n\{4\}/\varepsilon_n\{2\}$ versus Npart demonstrates this hierarchy more clearly and one even sees a convergence in $\varepsilon_n\{4\}/\varepsilon_n\{2\}$ at Npart less than Npart$\sim 20$.  

\begin{figure*}[h]
\centering
\begin{tabular}{c c}
\includegraphics[width=0.5\linewidth]{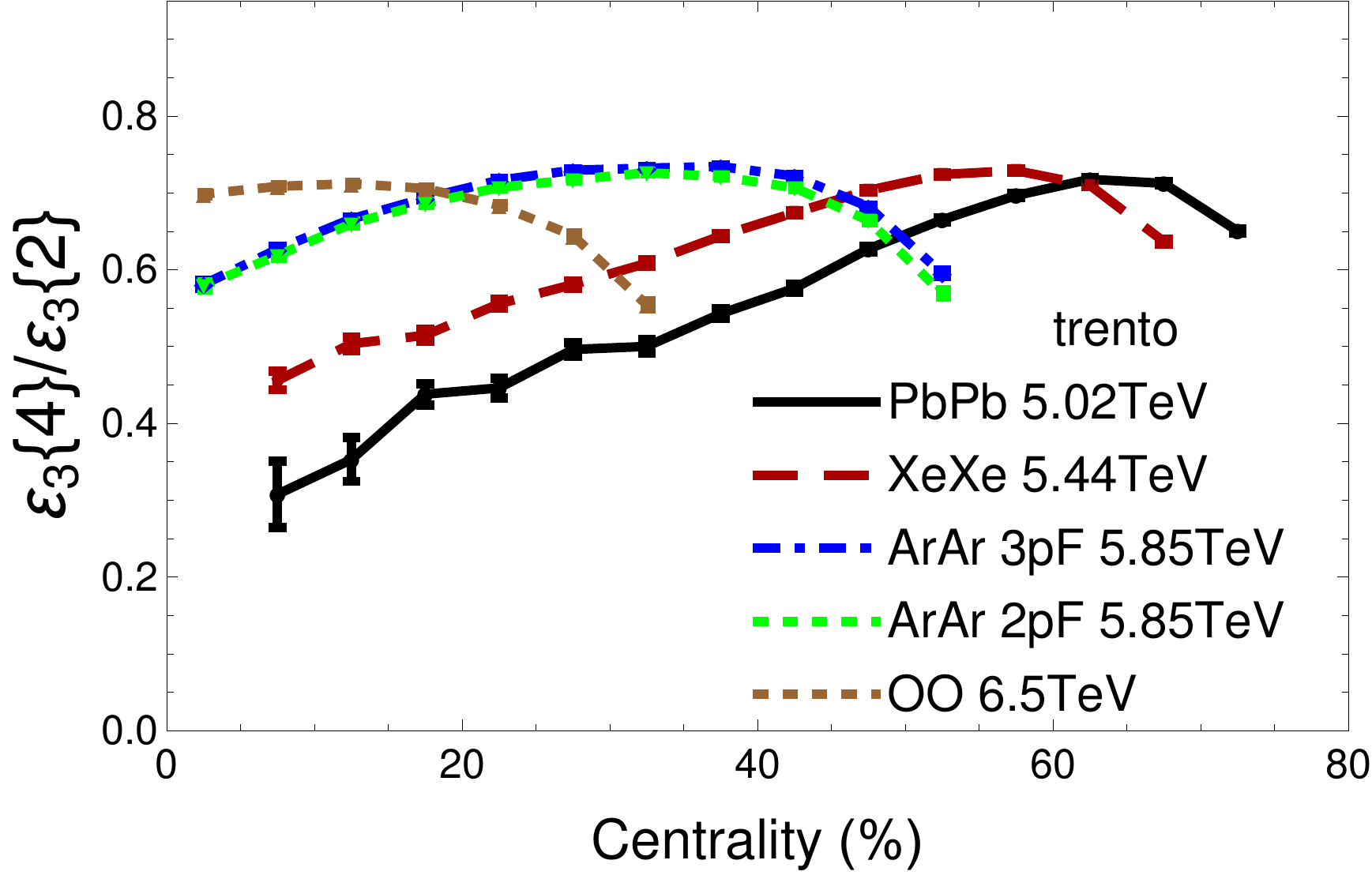} & \includegraphics[width=0.5\linewidth]{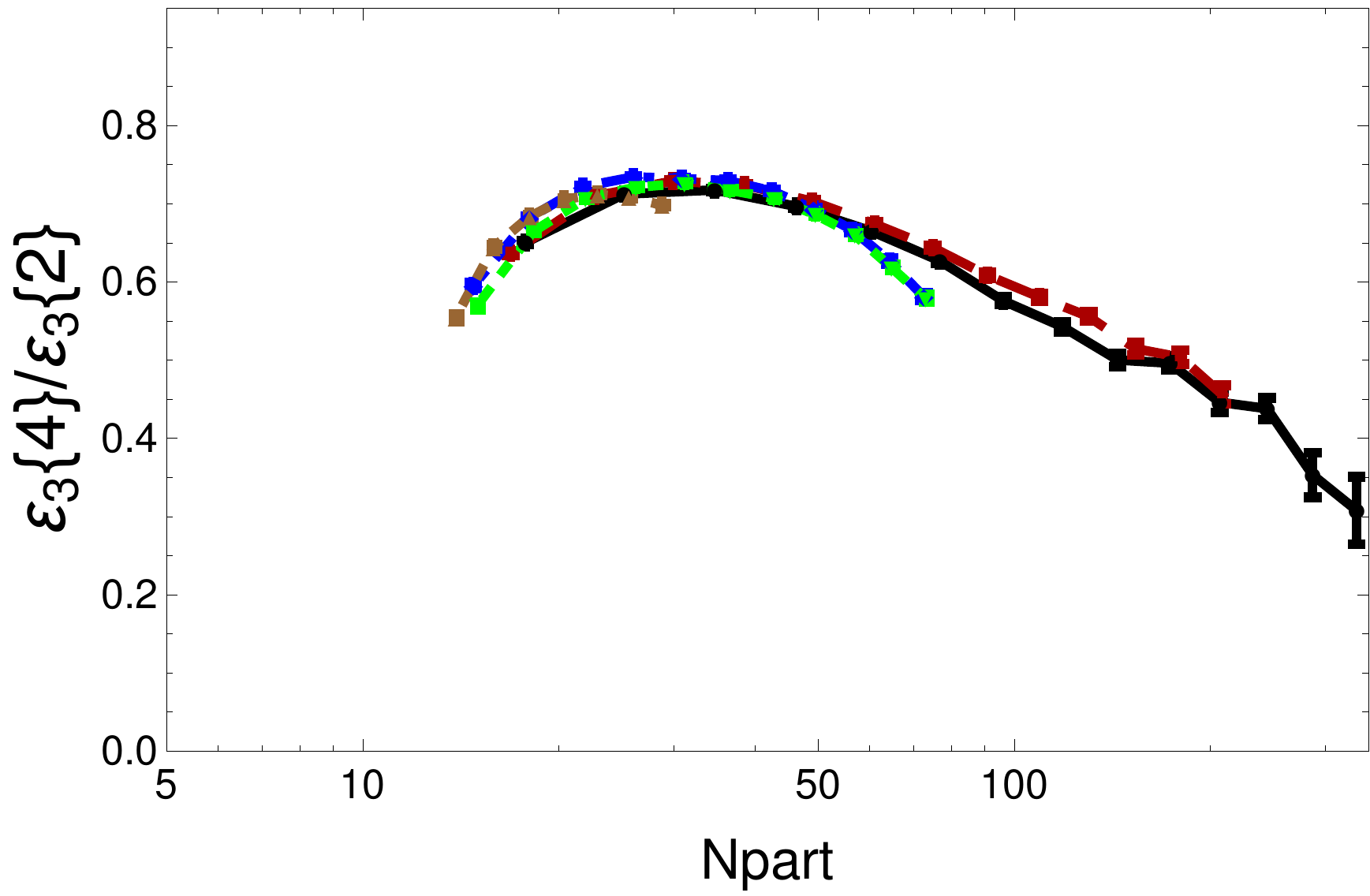}
\end{tabular}
\caption{(Color online) Ratio of $\varepsilon_3\{4\}/\varepsilon_3\{4\}$ for all system sizes plotted versus centrality (left) and Npart (right).  Note that 3pF denotes a 3 parameter Wood Saxon fit and 2pF denotes the typical two parameter Wood Saxon. }
\label{fig:cumu_v3}
\end{figure*}

A similar picture is seen for triangular eccentricity in Fig.\ \ref{fig:cumu_v3}, where the scaling by centrality is not very illuminating.  However, $\varepsilon_3\{4\}/\varepsilon_3\{2\}$ appears to collapse on a nearly universal curve when plotted by Npart.  This is an especially interesting result because it implies that if the triangular flow fluctuations are based on the same physics as one decreases the system size, then one expects them to collapse onto the same curve when plotted versus Npart. The only snag in this argument is if triangular flow fluctuations exhibit any non-linear response.  While this doesn't appear to be the case \cite{Giacalone:2017uqx}, we note that this is a very statistics hungry observable so it is quite difficult to test.

In Figs.\ \ref{fig:v2v3}-\ref{fig:cumu_v3} the  3pF signifies the 3 parameter Fermi distribution fit for the Wood-Saxon distribution in Eq. (\ref{e:3PF}) and 2pF signifies the 2 parameter Fermi distribution fit in Eq. (\ref{e:2PF}) are compared.  We find no visible difference between the two models and, therefore, use the 3pF in the rest of our calculations.

\subsection{Predictions from linear+cubic response}\label{sec:cubic}

In  \cite{Noronha-Hostler:2015dbi} it was found that linear response is insufficient to reproduce the flow fluctuations (specifically a large $\varepsilon_2$ produces an enhanced $v_2$ beyond what would be predicted from cubic response) in peripheral collisions.  The study was conducted for PbPb collisions at $\sqrt{s_{NN}}=2.76$ TeV and there was a centrality dependence in the non-linear response such that it was most relevant in peripheral collisions.  These results helped to explain the so called ``banana" plots from \cite{Niemi:2015qia}. 

\begin{figure*}[h]
\centering
\includegraphics[width=\linewidth]{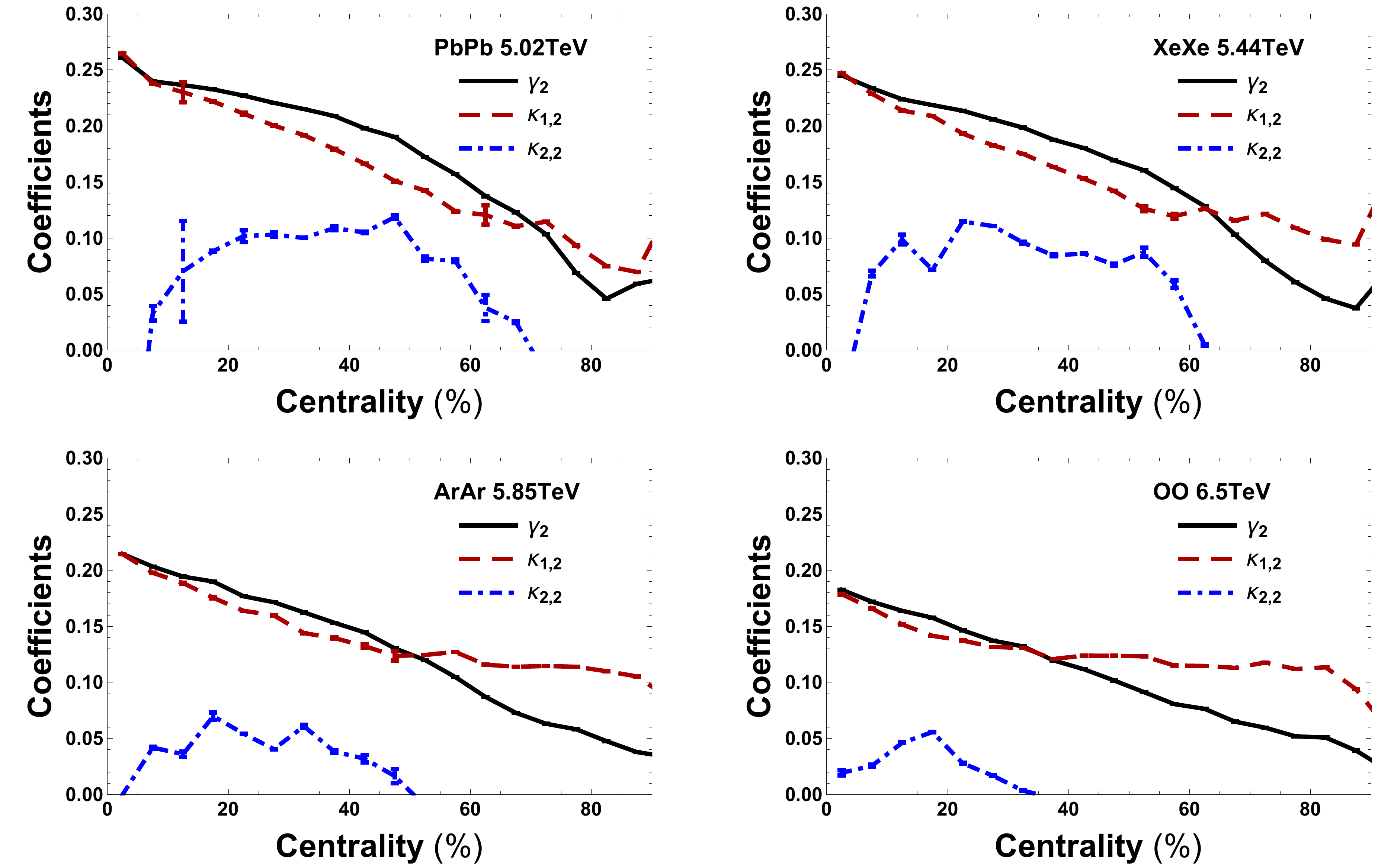} 
\caption{(Color online) Scaling coefficients extracted from Eq.\ (\ref{eqn:gamma1}) for linear only scaling, $\gamma_2$,and Eqs.\ (\ref{eqn:nonlinear}) for linear+cubic scaling, $\kappa_{1,2}$ and $\kappa_{2,2}$ are plotted versus centrality. All collisional systems are shown.}
\label{fig:kappas}
\end{figure*}
As of yet, the effects of linear vs. nonlinear response has not be studied versus system size.  Here we go beyond the linear response predictor for the final flow harmonics discussed in Eq.\ (\ref{eqn:linear}) and add in cubic response such that
\begin{equation}\label{eqn:precubic}
V_n=\kappa_{1,n} \mathcal{E}_n+\kappa_{2,n} |\varepsilon_n^2|\mathcal{E}_n
\end{equation}
where the coefficients $\kappa_{1,n}$ and $\kappa_{2,n}$ are obtained by minimizing the residual (see \cite{Noronha-Hostler:2015dbi}) using the Pearson coefficient. They are defined as:
\begin{eqnarray}\label{eqn:nonlinear}
\kappa_n&=&\frac{Re\left(\langle |\varepsilon_n|^6\rangle \langle v_n\varepsilon_n^*\rangle-\langle |\varepsilon_n|^4\rangle \langle v_n\varepsilon_n^*|\varepsilon_n|^2\rangle\right)}{\langle |\varepsilon_n|^6\rangle \langle |\varepsilon_n|^2\rangle-\langle |\varepsilon_n|^4\rangle^2}\\
\kappa_n^{\prime}&=&\frac{Re\left(-\langle |\varepsilon_n|^4\rangle \langle v_n\varepsilon_n^*\rangle+\langle |\varepsilon_n|^2\rangle \langle v_n\varepsilon_n^*|\varepsilon_n|^2\rangle\right)}{\langle |\varepsilon_n|^6\rangle \langle |\varepsilon_n|^2\rangle-\langle |\varepsilon_n|^4\rangle^2}
\end{eqnarray}
These coefficients are not positive definite and do become negative for very peripheral or very central collisions.  However, in peripheral collisions we find that the cubic response is only a good predictor when $\kappa_{2,n}>0$, as will be discussed below.

In Fig.\ \ref{fig:kappas} the linear coefficient, $\gamma_2$, is compared to the linear coefficient (when cubic response is present) $\kappa_{1,2}$ and  the cubic response coefficient $\kappa_{2,2}$.  Across all system sizes we find that $\gamma_2$ and $\kappa_{1,2}$ are fairly similar.  Generally, the linear coefficients decrease with increasing centrality.  The cubic response coefficient has a similar behavior as in \cite{Noronha-Hostler:2015dbi} in that it is the most relevant in PbPb collisions beyond $40\%$ centrality.  However, in that paper, only collisions up to $60\%$ centrality were considered.  Here we find that beyond $70\%$ centrality in PbPb collisions that $\kappa_{2,2}$ becomes negative.  

\begin{figure}[h]
\centering
\includegraphics[width=\linewidth]{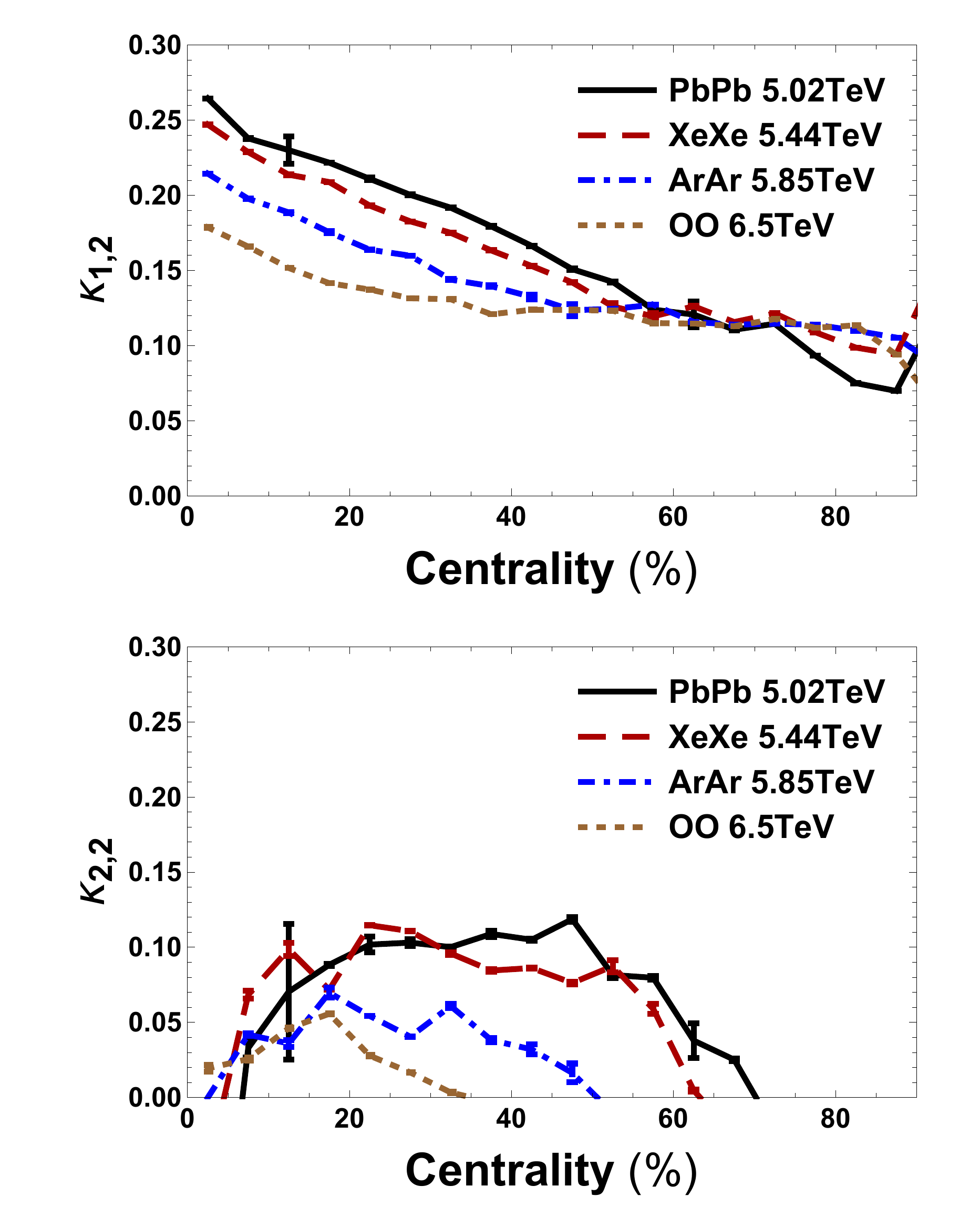} \caption{(Color online) Linear+cubic scaling coefficients extracted from Eqs.\ (\ref{eqn:nonlinear}), $\kappa_{1,2}$ (top) and $\kappa_{2,2}$ (bottom), are plotted versus centrality comparing the different collisional systems.}
\label{fig:kap}
\end{figure}

In Fig. \ref{fig:kap} these same coefficients are plotted versus the system size.  Here one can see more clearly that the linear response coefficients, $\kappa_{1,2}$, all decrease somewhat with system size (in central collisions only) and that the centrality dependence is smaller in small systems i.e. $\kappa_{1,2}$ becomes relatively flat across centrality.  In contrast, $\kappa_{2,2}$ decreases more dramatically as the system size is decreased and also for the smallest system of OO collisions there is only a small centrality window where $\kappa_{2,2}$ is positive.  

\begin{figure*}[h]
\centering
\includegraphics[width=\linewidth]{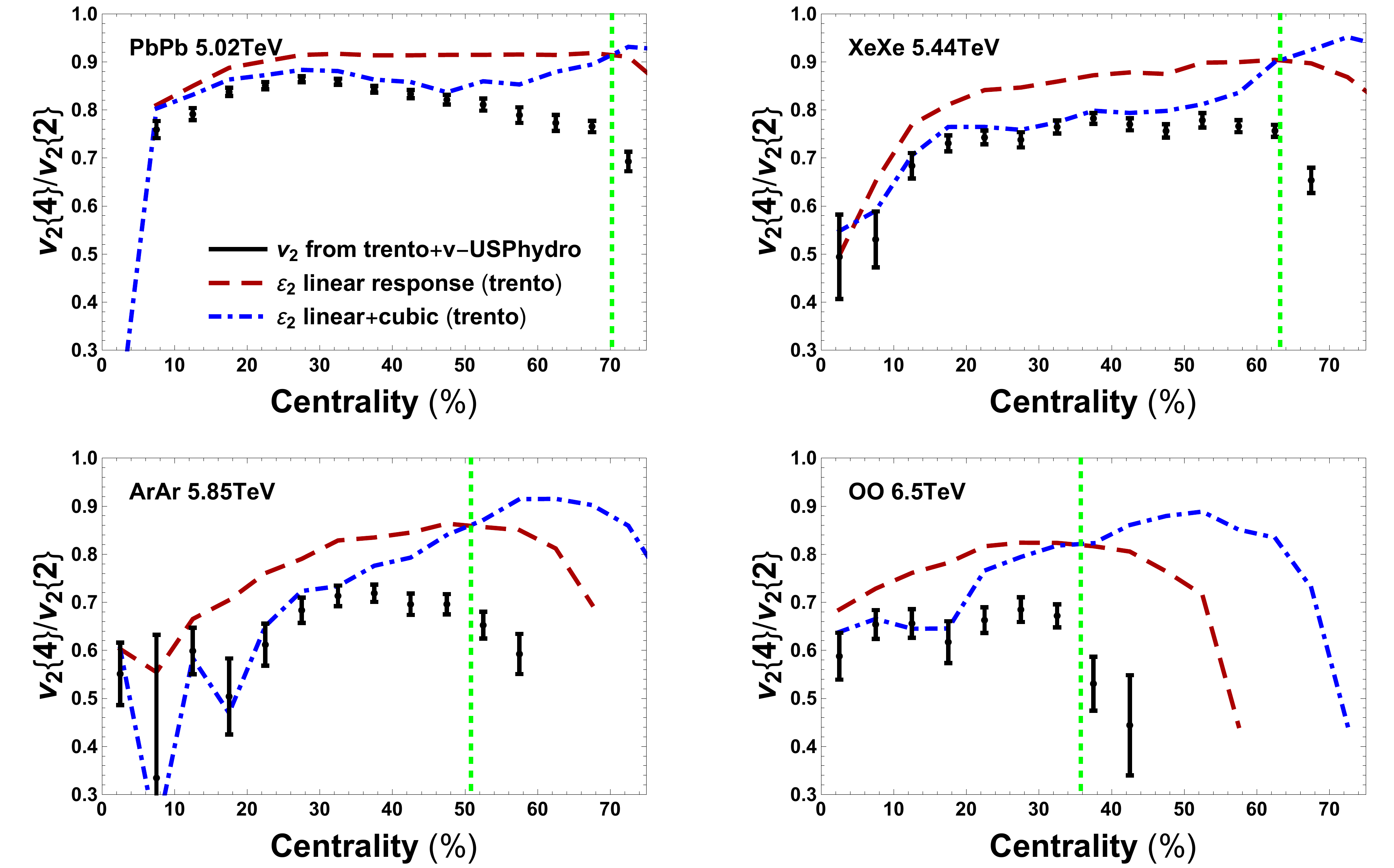} \caption{(Color online) Predictions for $v_2\{4\}/v_2\{2\}$ from linear scaling (red long dashed line) vs. linear+cubic scaling (blue dot dashed line).  The actual $v_2\{4\}/v_2\{2\}$  are shown in black with statistical error bars.  The green vertical line is the point where the cubic coefficient $\kappa_{2,2}$ becomes negative.}
\label{fig:predict}
\end{figure*}

The entire purpose of this exercise is to understand our ability to predict the final flow harmonics from the initial state alone.  Thus, we then use  Eq.\ \ref{eqn:linear} for linear response and Eq.\ \ref{eqn:precubic} for linear+cubic response to calculate the predicted $v_n$ in every event given only the eccentricity and then use this information to calculate the fluctuations using $v_2\{4\}/v_2\{2\}$.  The results of this exercise are shown in Fig.\ \ref{fig:predict}.  There the green dashed line indicates the point where $\kappa_{2,2}$ changes sign.  

Across all systems sizes we find that non-linear response is necessary to understand the observable $v_2\{4\}/v_2\{2\}$ and using only $\varepsilon_2\{4\}/\varepsilon_2\{2\}$ assuming only linear response will always over-predict the true result. Certainly, $\varepsilon_2\{4\}/\varepsilon_2\{2\}$ does provide a reasonable first order estimate but one should caution that this estimate works best in central collisions (as was discussed in \cite{Giacalone:2017uqx}). 

However, there is more to be learned from this plot.  For instance, the point where $\kappa_{2,2}\leq 0$ is also the same point where cubic response no longer is a better predict for $v_2\{4\}/v_2\{2\}$ than linear response. In very peripheral collisions ($>70\%$ centrality in PbPb collisions and $>35\%$ centrality in OO collisions), neither linear nor linear+cubic response adequately describe the final flow results.  One may be able to rectify this by higher order response terms
or this response could be arising from missed harmonic contributions.  We leave this exercise for a future work.  

Finally, we want to emphasize that in central to mid-central collisions the linear+cubic response is able to very accurately predict the final flow fluctuations.  For most centralities, the difference between the predicted flow fluctuations vs. the final flow harmonics is less than $<5\%$.  Only as one approaches a negative $\kappa_{2,2}\leq 0$, is a large deviation from the final flow harmonics seen.  Furthermore, the region where it is possible to make an accurate prediction for the final flow fluctuations shrinks as one shrinks the system size.  For oxygen 16, accurate flow fluctuations predictions using linear+cubic response are only possible between $0-20\%$ centrality.  This implies that for pPb collisions, linear+cubic response is likely only able to predict the flow fluctuations in very central collisions. 

\section{Results}\label{sec:results}

\begin{figure*}[h]
\centering
\includegraphics[width=\linewidth]{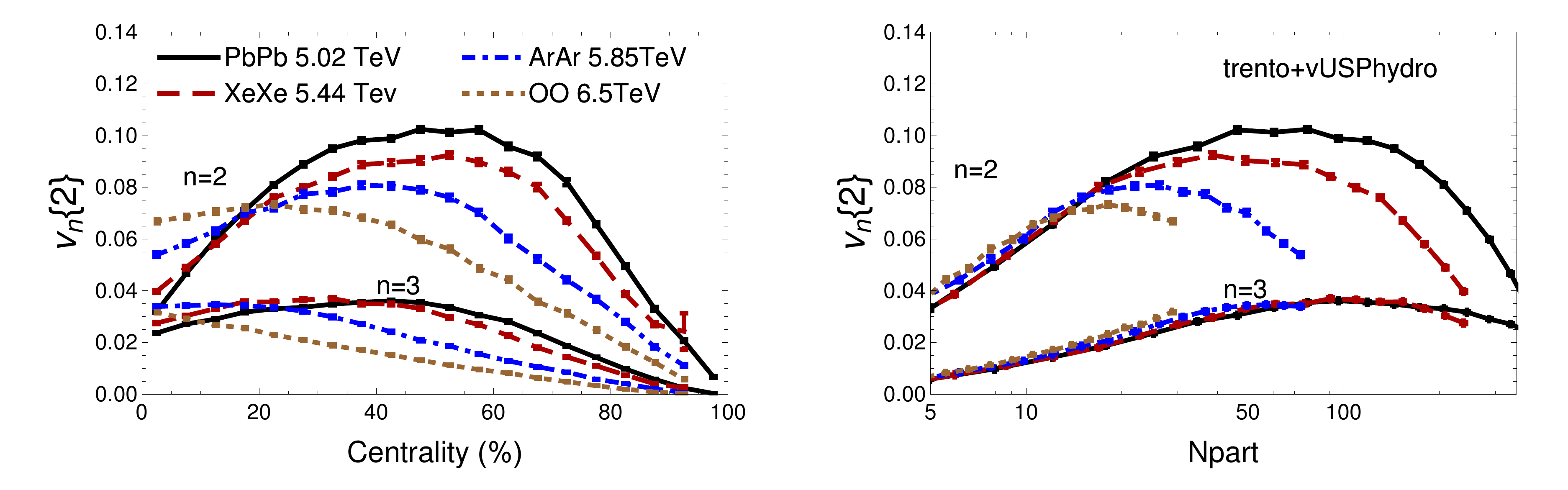}
\caption{(Color online) $v_n\{2\}$ for all collisional systems across centrality (left) and Npart (right).}
\label{fig:vn2}
\end{figure*}

In this section we cover all the results after running event-by-event relativistic viscous hydrodynamics across the different system sizes.  The most important flow observable that is typically used to constrain $\eta/s$ and other medium properties are integrated $v_n\{2\}$ calculations of elliptical and triangular flow.  Here we integrate all charged particles from $p_T=0.2-3$ GeV.  In Fig.\ \ref{fig:vn2} $v_2\{2\}$ and $v_3\{2\}$ across both centrality (left) and Npart (right). Comparing Fig.\ \ref{fig:eccrat} to Fig.\ \ref{fig:vn2}, we see very similar trends.  As the system size shrinks, $v_2\{2\}$ is more suppressed comparatively than $v_3\{2\}$, as expected from the eccentricities. Additionally, for smaller systems the peak broadens and moves towards more central collisions to the point that in OO collisions there is almost no peak left in $v_2\{2\}$ and it disappears entirely in $v_3\{2\}$. Finally, the sift in the peaks towards more central collisions does appear to shift more for triangular flow than for elliptical flow, as predicted from the eccentricities.

When the flow harmonics are plotted versus Npart, it appears that $v_3\{2\}$ falls an approximately universal curve.  Thus, regardless of system size, $v_3\{2\}$ appears to be entirely driven primarily by Npart. Even for $v_2\{2\}$ for $Npart<20$, $v_2\{2\}$ also appears to fall on a universal curve.  

\begin{figure*}[h]
\centering
\includegraphics[width=\linewidth]{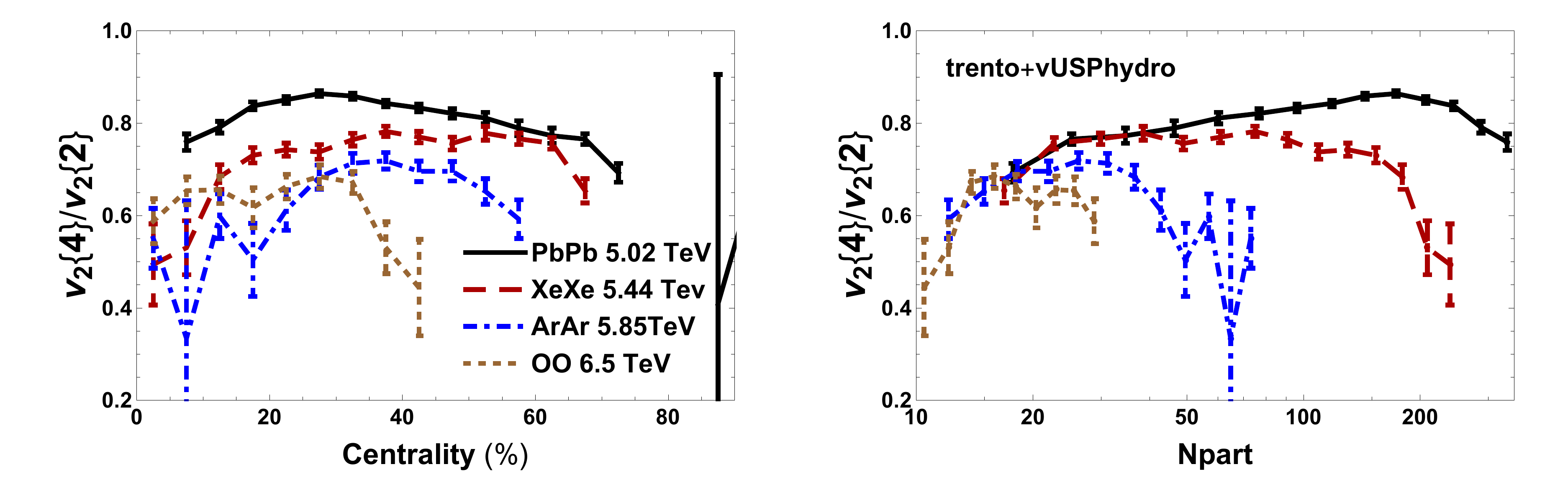}
\caption{(Color online) $v_2\{4\}/v_2\{2\}$ ratio for all system sizes plotted versus centrality (left) and Npart (right). }
\label{fig:trace}
\end{figure*}

It is possible to study the distribution of flow harmonics on an event-by-event basis using multi-particle cumulants.  The ratio of $v_n\{4\}/v_n\{2\}$ is smaller for a wider distribution of $v_n$'s and approaches 1 for a narrow distribution.  Just as for the ratio of the eccentricities of $\varepsilon_n\{4\}/\varepsilon_n\{2\}$ in Sec.\ \ref{sec:ec}, plotting versus centrality does not provide a very clear picture across system size.  However, plotting versus Npart the ratio $v_n\{4\}/v_n\{2\}$ converges for small enough Npart ($Npart<20$).  For larger Npart, the system size is inversely related to the size of the $v_2$ fluctuations.

While we are able to investigate the ratio $\varepsilon_3\{4\}/\varepsilon_3\{2\}$ easily in Sec.\ \ref{sec:ec}, we note that one requires very large statistics to reasonable calculate $v_3\{4\}/v_3\{2\}$ in hydrodynamics.  Using the $\sim 31,000$ events here, we obtain quite large statistical error bars in Fig. \ref{fig:trace}.  Plotting both compared to centrality and Npart it is difficult to draw any conclusions and we note that significantly more events would be needed before a reasonable comparison could be made.

\begin{figure*}[h]
\centering
\includegraphics[width=\linewidth]{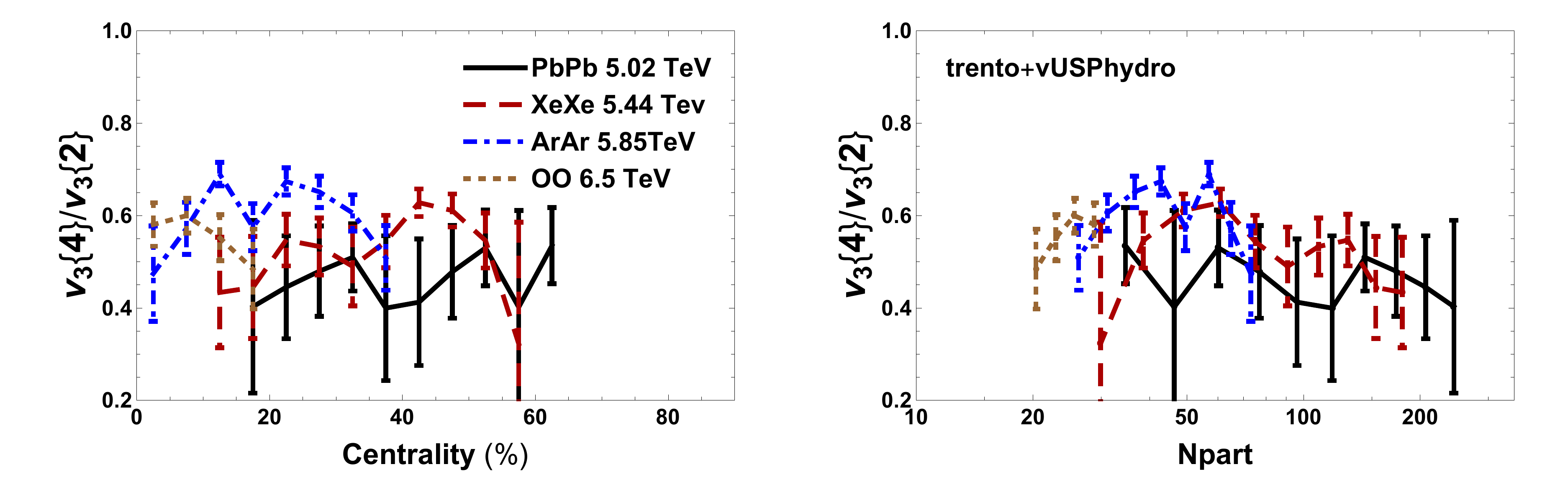}
\caption{(Color online) $v_3\{4\}/v_3\{2\}$ ratio for all system sizes plotted versus centrality (left) and Npart (right). }
\label{fig:trace}
\end{figure*}

In the follow sections we explore multiparticle cumulants of mixed harmonic observables: symmetric cumulants and event plane correlations. 
\subsection{Symmetric Cumulants}

Symmetric cumulants have been measured in small systems \cite{Sirunyan:2017uyl} where the degree that the elliptical flow fluctuates versus the triangular flow is measured by $NSC(3,2)$ (here we discuss normalized symmetric cummulants to remove effects from the magnitude of the flow harmonics). $NSC(3,2)$ has been found to scale with the multplicity such that pp, pPb, and PbPb collisions producing the same number of particles all converge to the same normalized symmetric cumulant. Initially, this was thought to be a sign of collective flow from a hydrodynamic-like picture.  However, recent CGC calculations have called this assumption into question \cite{Mace:2018vwq,Mace:2018yvl,Kovchegov:2018jun}. Thus, a natural question arises if one can see this convergence of the symmetric cumulants vs. multiplicity across a wide range of system sizes within a hydrodynamic framework.

On an event-by-event basis the different order flow harmonics not only fluctuate in strength but often have a non-trivial relationship between how they fluctuate compared to each other.  In order to determine this geometrical relationship the observable known as symmetric cumulants was suggested:
\begin{equation}\label{eqn:NSC}
NSC(m,n)=\frac{\langle v_m^2 v_n^2\rangle-\langle v_m^2\rangle\langle v_n^2\rangle}{\langle v_m^2\rangle\langle v_n^2\rangle},
\end{equation}
which has been studied extensively \cite{Bhalerao:2011yg,ALICE:2011ab,Zhou:2015eya,ALICE:2016kpq,Zhou:2016eiz,Zhu:2016puf,Gardim:2016nrr,Giacalone:2016afq,PhysRevC.95.044902,Ke:2016jrd,Eskola:2017imo}. It was found that $NSC(3,2)$ can be well described by its eccentricities alone whereas symmetric cumulants with high-order harmonics have non-linear behavior likely driven by viscous effects. Additionally, it has been shown that there is a connection between symmetric cumulants and event plane correlations \cite{Giacalone:2016afq}.  In this work all symmetric cummulants are calculated using multiplicity weighing and centrality rebinning \cite{Gardim:2016nrr}. 

\begin{figure*}[h]
\centering
\includegraphics[width=1\linewidth]{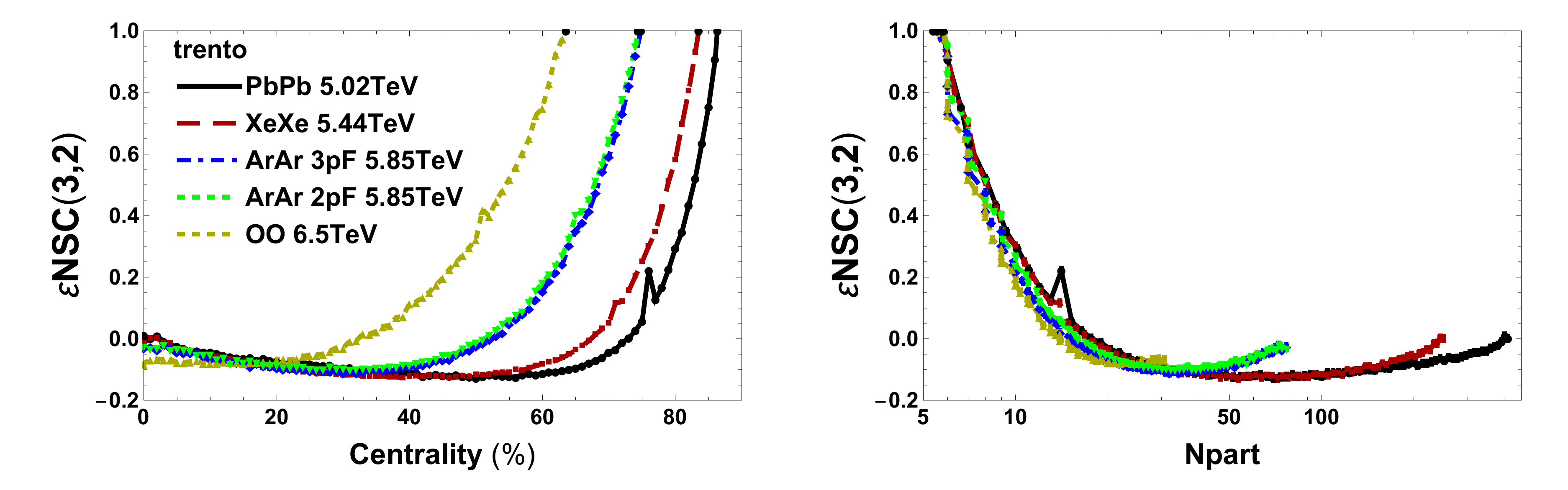}
\caption{(Color online) Normalized symmetric cumulants results of  $\varepsilon NSC(3,2)$ for the eccentricities in  PbPb $\sqrt{s_{NN}}=5.02$ TeV collisions, XeXe $\sqrt{s_{NN}}=5.44$ TeV collisions ArAr $\sqrt{s_{NN}}=5.85$ TeV collisions, OO $\sqrt{s_{NN}}=6.5$ TeV collisions scaled by centrality (left) and Npart(right).}
\label{fig:sc32ecc}
\end{figure*}

First we compare the scaling of the eccentricities, $\varepsilon NSC(3,2)$, in Fig.\ \ref{fig:sc32ecc} with system size. For the correlation between elliptical and triangular flow there is a clear hierarchy between smaller and larger systems where smaller systems demonstrate a larger correlation between $v_2$ and $v_3$. This correlation arises directly from the Npart as shown in Fig.\ \ref{fig:sc32ecc}  on the left.  All symmetric cumulants collapse onto a single curve except for very central collisions.  Here we only compare $\varepsilon NSC(3,2)$ since higher order harmonics such as $NSC(4,2)$ have non-linear effects that is dependent on other factors such as the viscosity. 

\begin{figure*}[h]
\centering
\includegraphics[width=1\linewidth]{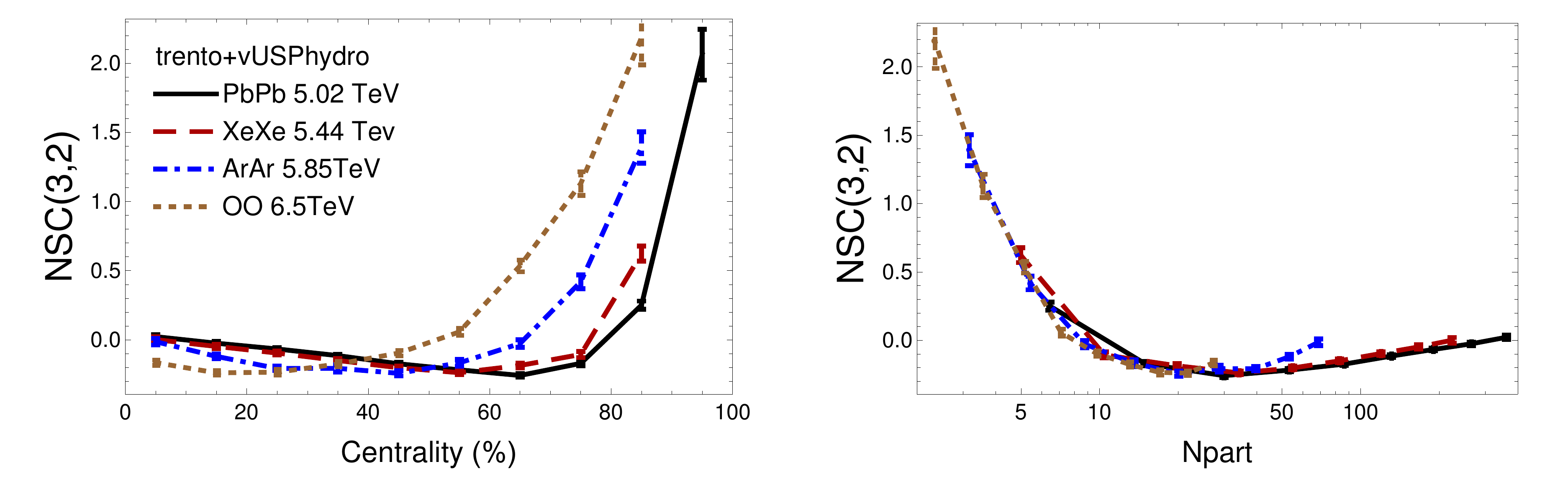}
\caption{(Color online) Normalized symmetric cumulants results of  $NSC(3,2)$ for all charged particles in  PbPb $\sqrt{s_{NN}}=5.02$ TeV collisions, XeXe $\sqrt{s_{NN}}=5.44$ TeV collisions ArAr $\sqrt{s_{NN}}=5.85$ TeV collisions, OO $\sqrt{s_{NN}}=6.5$ TeV collisions scaled by centrality (left) and Npart(right).}
\label{fig:sc32}
\end{figure*}

\begin{figure*}[h]
\centering
\includegraphics[width=1\linewidth]{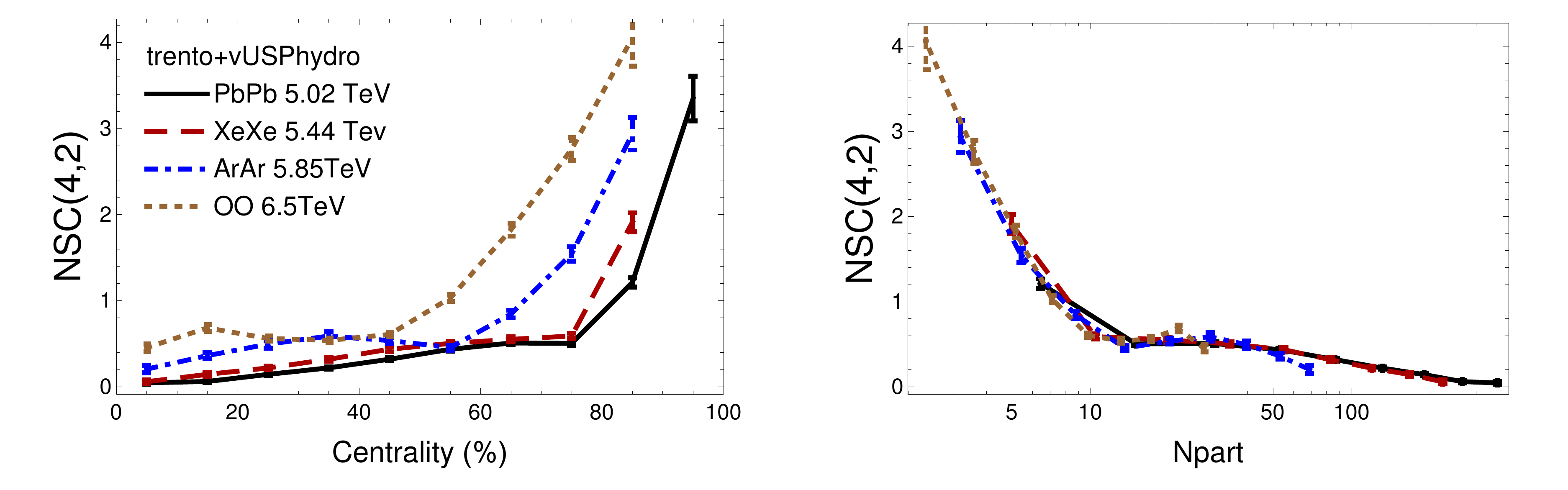}
\caption{(Color online) Normalized symmetric cumulants results of  $NSC(4,2)$ for all charged particles in  PbPb $\sqrt{s_{NN}}=5.02$ TeV collisions, XeXe $\sqrt{s_{NN}}=5.44$ TeV collisions ArAr $\sqrt{s_{NN}}=5.85$ TeV collisions, OO $\sqrt{s_{NN}}=6.5$ TeV collisions scaled by centrality (left) and Npart(right).}
\label{fig:sc42}
\end{figure*}

\begin{figure*}[h]
\centering
\includegraphics[width=1\linewidth]{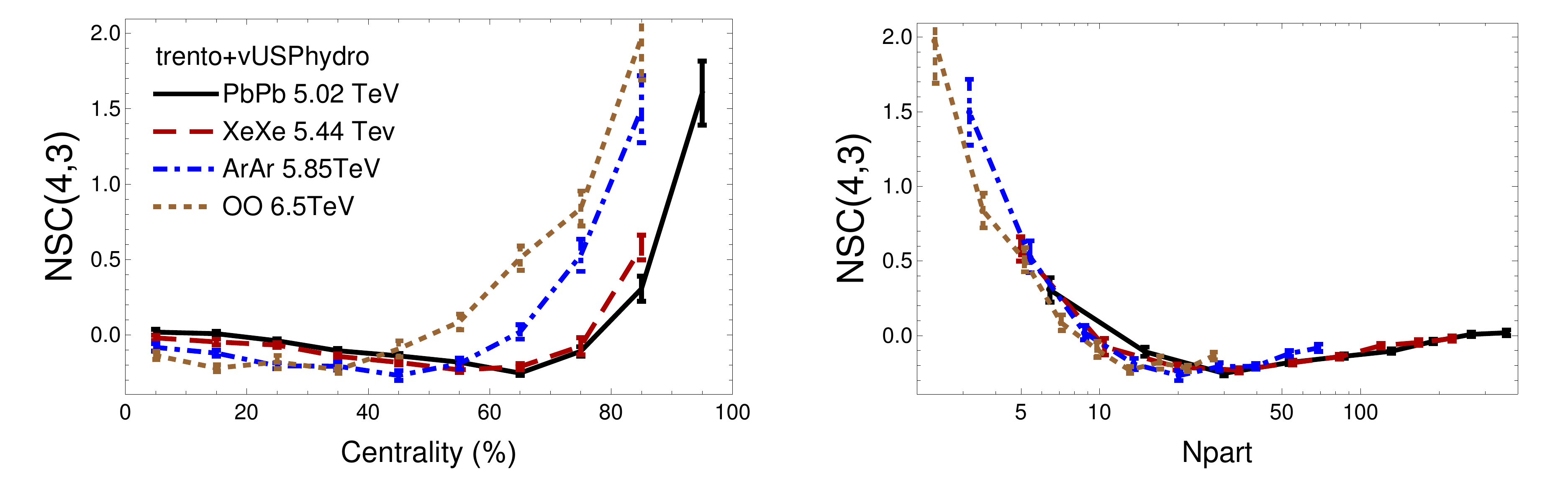}
\caption{(Color online) Normalized symmetric cumulants results of  $NSC(4,3)$ for all charged particles in  PbPb $\sqrt{s_{NN}}=5.02$ TeV collisions, XeXe $\sqrt{s_{NN}}=5.44$ TeV collisions ArAr $\sqrt{s_{NN}}=5.85$ TeV collisions, OO $\sqrt{s_{NN}}=6.5$ TeV collisions scaled by centrality (left) and Npart(right).}
\label{fig:sc43}
\end{figure*}

In \cite{Sirunyan:2017uyl} it was found that when one scales by the multiplicity that the normalized symmetric cumulants of different system sizes collapse onto a single curve. One should note, however, that non-flow contributions should be carefully taken into account   \cite{Huo:2017nms}.  Finally, work has shown that in small systems symmetric cumulants can provide information about proton substructure \cite{Albacete:2017ajt}. Because of these previous results we study the system size effects of  $NSC(m,n)$ with a special interest in their scaling behavior with Npart.  In Figs.\ \ref{fig:sc32}-\ref{fig:sc43} the symmetric cumulants $NSC(3,2)$, $NSC(4,2)$, and $NSC(4,3)$ are shown, respectively. The results are scaled by both centrality and Npart.

When plotted versus centrality we find all  $NSC(m,n)$ show a monotonically increasing behavior of $NSC(m,n)$ for peripheral collisions (centralities $\gtrapprox 50\%$) as the system size is decreased.  More central collisions (centralities of $\lessapprox 40\%$) have relatively similar results regardless of the system size, although in all cases the absolute value of $|NSC(m,n)|$ is closest to zero for the largest system size of PbPb collisions. 

Once $NSC(m,n)$ is scaled by Npart (used as a proxy for multiplicity), we find that our results appear to collapse onto a universal curve regardless of system size.  This is consistent with our eccentricity results for $\varepsilon NSC(m,n)$ as shown in Fig.\ \ref{fig:sc32ecc}.  Although, from the eccentricity results one would expect slight deviations from this curve for the most central collisions of each system.  Thus, for $NSC(m,n)$ it does appear that the symmetric cumulants deviate from the eccentricities for small Npart.  Only a small deviation is seen for the most central ArAr collision but all other systems appear to fit to the universal curve even for central collisions. 

\section{Event Plane Correlations}

Most flow observables focus on the magnitude of the flow vectors (e.g. symmetric cumulants and multi-particle cumulants).  However, since flow harmonics are, in fact, vectors and carry information about both the magnitude and the angle of the flow, ATLAS first measured event-plane correlations in \cite{Aad:2014fla} to understand how these angles fluctuation on an event by event basis.  However, these measurements also carry non-trivial information about the magnitude of the flow harmonics as well and can be even directly correlated to the symmetric cumulants  \cite{Giacalone:2016afq}.  That being said, event plane correlations can help to distinguish different viscosities \cite{Niemi:2015qia} and do appear to be somewhat sensitive to the equation of state \cite{Noronha-Hostler:2018zxc}. 

Recently at STAR \cite{Adamczyk:2017byf} the event plane correlations were measured using
\begin{equation}
C_{n,m,n+m}=\langle v_m v_n v_{m+n} \cos \left(m\Psi_m+n\Psi_n-(m+n)\Psi_{m+n}\right) \rangle,
\end{equation}
where it is not normalized by the magnitude of the flow harmonics.  Here we compare $C_{n,m,n+m}$ because of its sensitivity to medium properties such that it could be an important benchmark in constraining transport coefficients across system size.

\begin{figure*}[h]
\centering
\includegraphics[width=1\linewidth]{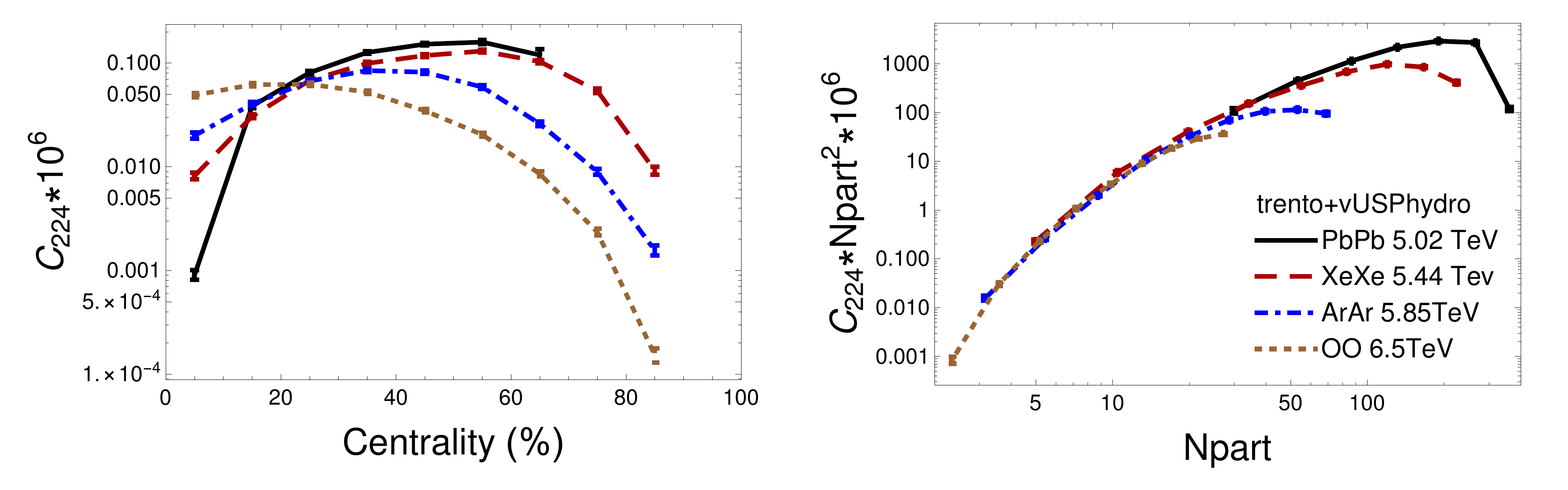}
\caption{(Color online) $C_{224}*10^6$ for all system sizes plotted versus centrality (left) and Npart. }
\label{fig:ep224}
\end{figure*}

\begin{figure*}[h]
\centering
\includegraphics[width=1\linewidth]{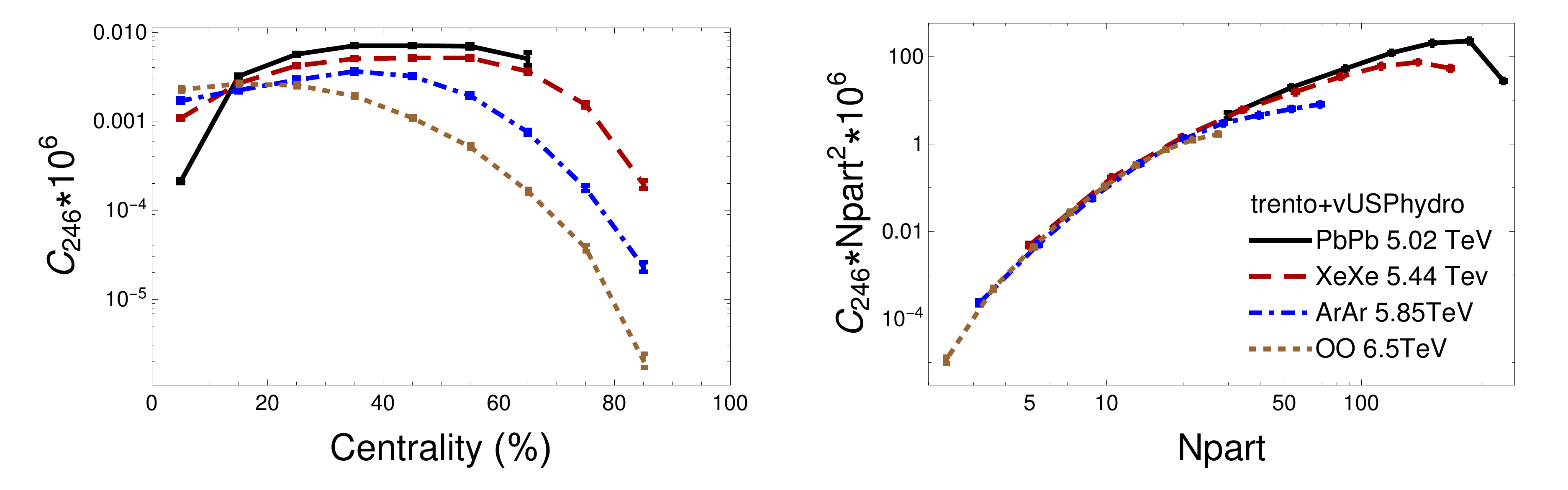}
\caption{(Color online) $C_{235}*10^6$ for all system sizes plotted versus centrality (left) and Npart. }
\label{fig:ep246}
\end{figure*}

\begin{figure*}[h]
\centering
\includegraphics[width=1\linewidth]{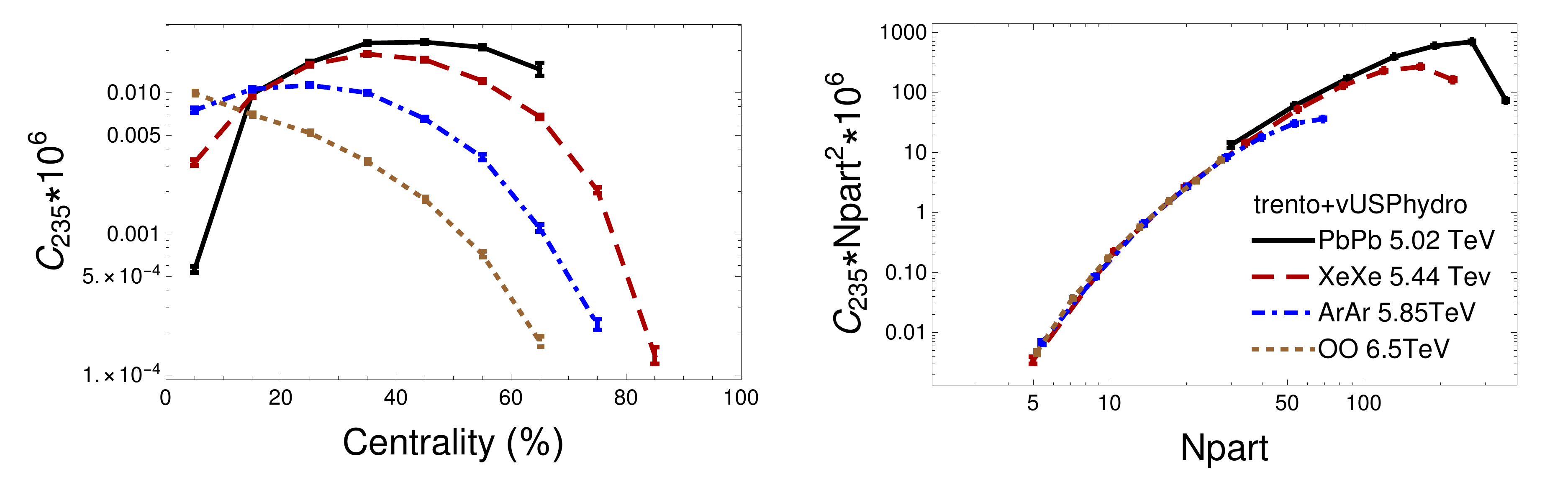}
\caption{(Color online)  $C_{246}*10^6$ for all system sizes plotted versus centrality (left) and Npart.}
\label{fig:ep235}
\end{figure*}

In Figs.\ \ref{fig:ep224}-\ref{fig:ep235} we compare $C_{224}$, $C_{246}$, and $C_{235}$, respectively. Generally, we find that for very central collisions the smallest system has the largest $C_{n,m,n+m}$ but collisions with a centrality of $>20\%$ the inverse occurs and $C_{n,m,n+m}$ is the largest for the largest systems.  When we rescale by Npart and plot instead $C_{n,m,n+m}*Npart^2$ there is a hierarchy in central collisions where the largest system always has the largest $C_{n,m,n+m}*Npart^2$ but peripheral collisions converge to a universal curve.  The behavior looks nearly identical for all variations of $C_{n,m,n+m}*Npart^2$ when plotted versus Npart.  It would be interesting to vary medium properties such as the EOS, $\eta/s(T)$ and $\zeta/s(T)$ to see how the dependence of medium properties scales with the system size.  However, we leave this to a future work. 

\section{Conclusions}\label{sec:conclu}

In this paper we studied the system size dependence of flow observables in the soft sector at the purposed future LHC collisions of ArAr $\sqrt{s_{NN}}=5.85$ TeV and OO $\sqrt{s_{NN}}=6.5$ TeV compared to the previous ran collisions of PbPb $\sqrt{s_{NN}}=5.02$ TeV and XeXe $\sqrt{s_{NN}}=5.44$ TeV.  

The largest system is PbPb and the smallest system is OO.  At the maximum size for OO collisions i.e. central collisions, the radius is $R\sim 4 fm$ with a multiplicity of $M\sim 556$.  Comparing the same multiplicity in PbPb collision, the radius is over $60\%$ larger than OO collisions. A central OO collisions has relatively small eccentricities and is quite a bit hotter than a PbPb collisions with the same multiplicity, which will generally be cooler with a more dominant elliptical shape. That is evident when one scales $v_2\{2\}$ be Npart that demonstrates a clear hierarchy by system sizes (the largest systems produces the largest $v_2\{2\}$ in central to mid-central collisions).  In contrast, $v_3\{2\}$ falls on a universal curve regardless of system size.  

One of the more interesting results that we found was that as the system size decreases there is generally a more linear mapping between the initial eccentricities (both for elliptical and triangular flow) onto the final flow harmonics.  We find that cubic response is most relevant in the largest of systems and can provide an accurate prediction of $v_2\{4\}/v_2\{2\}$ for central to mid-central collisions but that for very peripheral collisions both linear and cubic response fail to predict the flow harmonics.  In fact, the centrality window of predictability shrinks with the system size such that for OO collisions we can only accurate predict $v_2\{4\}/v_2\{2\}$ from $0-20\%$ centrality whereas in PbPb collisions we can reasonable predict $v_2\{4\}/v_2\{2\}$ from $0-50\%$. For more peripheral collisions with $Npart\leq 20$ we find that $v_2\{4\}/v_2\{2\}$ collapses onto a universal curve, which is exactly where our predictions for linear or linear+cubic response fails.  While linear response can obtain the qualitative behavior of $v_2\{4\}/v_2\{2\}$ in these very peripheral collisions, it generally significantly over-predicts the ratio compared to the final flow harmonics after running hydrodynamics.

Here we made predictions for $v_2\{2\}$, $v_3\{2\}$, $v_2\{4\}/v_2\{2\}$, $v_3\{4\}/v_3\{2\}$, the symmetric cumulants $NSC(3,2)$, $NSC(4,2)$, $NSC(4,3)$  and the event plane correlations $C_{224}$, $C_{246}$, and $C_{235}$ and plot versus both the centrality and Npart. Generally, we found that scaling by Npart either would collapse all the results into a single universal curve or a clear hierarchy related to the system size is seen.  A universal curve seen  for the follow observables: $v_3\{2\}$, all symmetric cumulants, and possibly also $v_3\{4\}/v_3\{2\}$ (the eccentricities indicated that plotting versus Npart should have them fall onto a universal curve but our statistical uncertainties after running hydrodynamics are still too large to say for certain).  A hierarchy is seen in the observables: $v_2\{2\}$, $v_2\{4\}/v_2\{2\}$, $\langle p_T\rangle$, and all the the event plane correlations. We also found that for $\langle p_T\rangle$ there was not a strong dependence on the type of particle when it came to the scaling behavior.  Thus, we generally expect that all particles that flow would see the same suppression in $\langle p_T\rangle$ across centrality as the system size shrinks.

Due to the large gradients in small systems, it is not clear that hydrodynamics has the same transport coefficients across all system sizes \cite{Romatschke:2017vte,Heller:2015dha,Denicol:2017lxn,Strickland:2017kux,Blaizot:2017ucy,Behtash:2018moe}.  One way to explore this notion would be to compare observables that are known to be sensitive to medium properties across system size.  For this reason, we would encourage experimentalists to measure the event plane correlations, $C_{n,m,n+m}$,  across different system sizes.  Our predictions find a reasonable scaling with the system size but it would be interesting to investigate the sensitive to other medium properties as one changes the system size.  This is an exercise that we leave for a future work. 

In conclusion, we find that a system size scan at the LHC would be a very useful exercise to nail down the limits of relativistic hydrodynamics.  We would encourage experimentalists to measure not only the standard bread and butter observables such as multiparticle cumulants but also to investigate the scaling relationship for symmetric cumulants and event plane correlations.  From previous work, we have found that our best fitting results for PbPb collisions scale well to the smaller system size of XeXe collisions.  Thus, if our current framework of TRENTO initial conditions combined with relativistic hydrodynamics is still valid in small systems, we expect to see that our predictions are accurate for ArAr and OO collisions.  However, there is also the possibility that either medium properties change due to the presence of large gradients in small systems or that the physical assumptions behind the initial conditions must change already for OO collisions, thus, it is important to have a systematic system size scan at the LHC.

\section*{Acknowledgements}
The authors would like to thank Soumya Mohapatra for discussions related to this work. 
J.N.H. acknowledges the support of the Alfred P. Sloan Foundation and the Office of Advanced Research Computing (OARC) at Rutgers, The State University of New Jersey for providing access to the Amarel cluster and associated research computing resources that have contributed to the results reported here.

\section*{References}
\bibliography{allv3}

\end{document}